\documentclass[reprint, 
nofootinbib,
 amsmath,amssymb,
 aps, 
 prx,
]{revtex4-2}

\usepackage{graphicx}
\usepackage{dcolumn}
\usepackage{bm}

\usepackage{subcaption}
\usepackage{cleveref}
\usepackage{csquotes}

\usepackage{xcolor}

\begin{document}

\preprint{APS/123-QED}

\title{Learning to crawl: benefits and limits of centralized \emph{vs} distributed control}

\author{Luca Gagliardi}
\email{Contact author: luca.gagliardi@edu.unige.it}
\author{Agnese Seminara}
\email{Contact author: agnese.seminara@unige.it}
\affiliation{Machine Learning Genoa Center and Dept of Civil Chemical and Environmental Engineering, University of Genoa, villa Cambiaso, via Montallegro 1, 16145,  Italy}



\begin{abstract}
\noindent 
We present a model of a crawler
consisting of several suction units distributed along a straight line and connected by springs. 
The suction units are  rudimentary proprioceptors-actuators, which sense binary states of compression \emph{vs} elongation of the springs, and can either adhere or remain idle. 
Muscular contraction is not controlled by the crawler, but follows an endogenous, stereotyped wave. The crawler is tasked to learn patterns of adhesion that generate thrust in response to the wave of contraction.
Using tabular Q-learning we demonstrate that crawling can be learned by trial and error and we ask what are the benefits and limitations of distributed \emph{vs} centralized learning architectures. 
We find that by centralizing proprioceptive feedback and control, the crawler leverages long range correlations in the dynamics and ride the endogenous wave smoothly. The ensuing benefits are measured in terms of both speed and robustness to failure, although they come at increased computational cost. At the opposite extreme, purely distributed feedback and control  only leverages local information and yield a jerkier and slower crawling, although computationally cheap. Intermediate levels of centralization can negotiate fast and robust crawling while avoiding excessive computational burden, demonstrating the computational benefits of a hierarchical organization of crawling.
Our model unveils the trade-offs between crawling speed, robustness to failure, computational cost and information exchange that may shape biological solutions for crawling and could inspire the design of robotic crawlers.  
\end{abstract}

\maketitle


\section{\label{sec:intro}Introduction}

All organisms with a distributed nervous system need to coordinate their sensorimotor loop without descending commands from a central nervous system, typical of bilateral animals.
Despite the lack of centralized control, many distributed nervous systems reliably manage effective locomotion involving the synchronization of many body parts.

These nervous systems showcase different architectures reflecting varying degrees of distribution \emph{vs} centralization. In cnidarians, the nervous system consists only of diffuse nerve nets and is the most primitive of all animals'. Despite the lack of centralization, cnidaria are capable of complex behaviors. For example, \emph{Hydra vulgaris} achieves highly non-trivial coordination of its body parts to perform somersaults~\cite{Frost2023}. Furthermore, associative learning in the form of operant conditioning can be induced by either visual or mechanical stimuli for the box jellyfish~\emph{T.~Cystophora}~\cite{Bielecki2023}.

Echinoderms, such as starfish, also have a simple radial nervous system consisting of a nerve net of neurons. But while they have no brain, their nervous system is more structured than in the cnidarians, as the nerves connect to a central nerve ring~\cite{Freas2022}. Starfishes crawl efficiently using hundreds of independent tubular feet which generate robust locomotion without central control. The nerve ring establishes the direction of motion by integrating sensory information detected by the arms~\cite{Kano2019,Clark2018}. Given the direction of motion, coordination of the tube feet can emerge purely from a passive mechanical coupling, as showed with a minimal model where behavior of the tube feet is prescribed and no information propagates along the nerve net~\cite{Po2024,Heydari2020}.

Cephalopods, such as octopus, squid and cuttlefish, have evolved a complex nervous system whose size is comparable to that of vertebrates and endows them with impressive cognitive capabilities. They do have a central brain; however most of their nervous system is distributed throughout their arms which are covered in hundreds of suckers that can sense the environment as well as adhere to substrates. Like all soft-bodied animals, these organisms have to coordinate many degrees of freedom to achieve locomotion. For example, octopuses translocate on surfaces by leveraging their taste-by-touch sense, whereby suckers adhere to surfaces and sense them chemically upon contact~\cite{Allard2023}. The sensory information acquired by the periphery guides suckers' adhesion, yielding displacement of the organism along the surface. However, to what extent locomotion is controlled by the central nervous system \emph{vs} by the ganglia distributed along the arms is not known~\cite{Mather2017,Levy2015}. Intriguingly, coordination {in octopus} may or may not depend on precise proprioceptive feedback, since the molecular mechanisms for proprioception { and to what extent the brain stores positional information} is not understood~\cite{Gutnick2011,Zullo2009}. 

This question is part of a large body of literature {in neuromechanics} that broadly asks how do animals and robots control {various kinds of distributed locomotion} to achieve specific goals including energy efficiency, maneuverability, stability, speed (see  e.g.~\cite{dickinson2000,nishikawa2007neuromechanics} and references therein). { Approaches that range from a more centralized~\cite{ijspeert2008central} to a purely distributed control~\cite{Umedachi2010} have been put forward for various locomotion mechanisms in bio-inspired robotics, reviewed e.g.~in \cite{spagna2007distributed}.} 
Whether centralization increases locomotion speed is an open question 
(reviewed e.g.~in~\cite{neveln2019information}). On the one hand, centralization likely requires additional processing which may be too slow to be implemented at fast speed. On the other hand, the need for added stability at high speed may require centralized control. { The combination of CPG and local feedback was found beneficial for adaptation and versatility for amphibious centipedes~\cite{yasui2019decoding}, undulatory patterns in swimming and multi-leg terrestrial locomotion~\cite{yasui2025multisensory} and  quadruped locomotion~\cite{suzuki2021spontaneous}.}

{ Here we focus on adhesion-driven crawling:} we draw inspiration from {octopus} and focus on a toy model to ask what are the pros and cons of centralization \emph{vs} distribution. The crawler is modeled as a collection of suckers distributed along a straight line and connected by springs, and endowed with a rudimentary form of proprioception (extension \emph{vs} elongation of the springs) and a rudimentary form of control of the suckers (adhere \emph{vs} not adhere). Contraction of the springs is not controlled by the crawler but rather follows a stereotyped wave established by a central pattern generator (CPG). If suckers were to adhere at random times, the crawler center of mass would be unable to move on average: is it possible to learn a pattern of adhesion that enables net translocation, {purely from trial and error? \\

Learning is an underexplored aspect of crawling that is relevant to develop robotic crawlers in unpredictable environments, for example on irregular terrain or under  other environmental perturbations. In these cases, programming the robot to crawl is unfeasible and learning from trial and error becomes a viable option. \\
} 

To answer this question we develop a simulator of the crawler
and let it learn to translocate by trial and error using the Q-learning Reinforcement Learning algorithm. 
{We take the approach of fixing a widely used model of coupled oscillators, or template~\cite{Full}, where each unit has the ability to adhere. The mathematical model is similar to a large body of literature on the neuromechanics of crawling largely developed for larvae, worms and insects (e.g.~\cite{ayali2015comparative,koditschek2004mechanical,pearson2006assessing} and references within). Our crawler does not control muscle contraction, but only adhesion with the substrate. 
A main difference with most literature (but see~\cite{Mishra2020}) is that we do not program the crawler, but let it learn through the use of reinforcement learning, whereby we set the goal of translocating and the agent learns to crawl purely from trial and error. This approach allows us to assess how easy it is for a learning architecture to develop an efficient crawling, and how computationally demanding learning is. Efficiency is defined as the combination of fast translocation and robustness to failure of individual suckers.
Alternatively, one may target a specific dynamics and learn how to formulate a mathematical model that reproduces a specific attractor in the phase space~\cite{Ijspeert2013}. We ask whether the sole control of adhesion is sufficient to learn crawling with a distributed \emph{vs} centralized architecture. 
Our results complement previous literature considering patterns of contraction --rather than adhesion-- in fully centralized learning of crawling~\cite{Mishra2020} and fully distributed learning of swimming~\cite{Hartl2025}, as well as works where crawling emerges from prescribed control patterns, rather than learning (see e.g.~\cite{Tanaka2012,Po2024}), and where the redundancies of preprogrammed gait design lead alone to locomotion robustness, for instance over rough terrains \cite{Chong2023}.}\\
We compare the learnability, performance and robustness of centralized \emph{vs} distributed learning architectures. 
Despite the rudimentary proprioception and control, we find that 
the sole existence of a common reward --here the speed of the center of mass-- is sufficient for independent suckers to learn crawling. 
When all suckers are simultaneously controlled by a single agent, crawling is both faster and more robust to failure, partially overcoming the crawler's poor sensorimotor ability.
However, the computational cost of the central control scales exponentially with the number of suckers and springs.
We then introduce a shallow hierarchical architecture where control is centralized in few control centers. We find that this partial centralization achieves nearly optimal performance and robustness with a limited increase in computational burden. 

{In our setting, crawling is achieved by \enquote{surfing} the wave generated by the central pattern generator. The existence of a central pattern generator is commonly assumed and has been demonstrated in various 
systems (see e.g.~\cite{cpg_data}). Note however that in the absence of a central pattern generator, crawling may still be achieved by proprioceptive feedback alone \cite{WEN2012750,boyleetal_2012,cengiz_paoletti_maha_2016}, or by leveraging synchronization of motor control with some natural frequency of the system (for example in the context of human inspired biped locomotion via imitation learning~\cite{imitation_learning_biped}, or limb oscillation via experimental analysis~\cite{Hatsopoulos01031996}, or for flapping wings in flight~\cite{lynch}).}


We hope that our work will serve as a benchmark to explore the optimal level of centralization in more realistic scenarios for natural and artificial crawling, where constraints on speed, robustness and computational costs will all concur to define optimality.

\section{\label{sec:methods}Model and learning}
\begin{figure}
    \includegraphics[width=\linewidth]{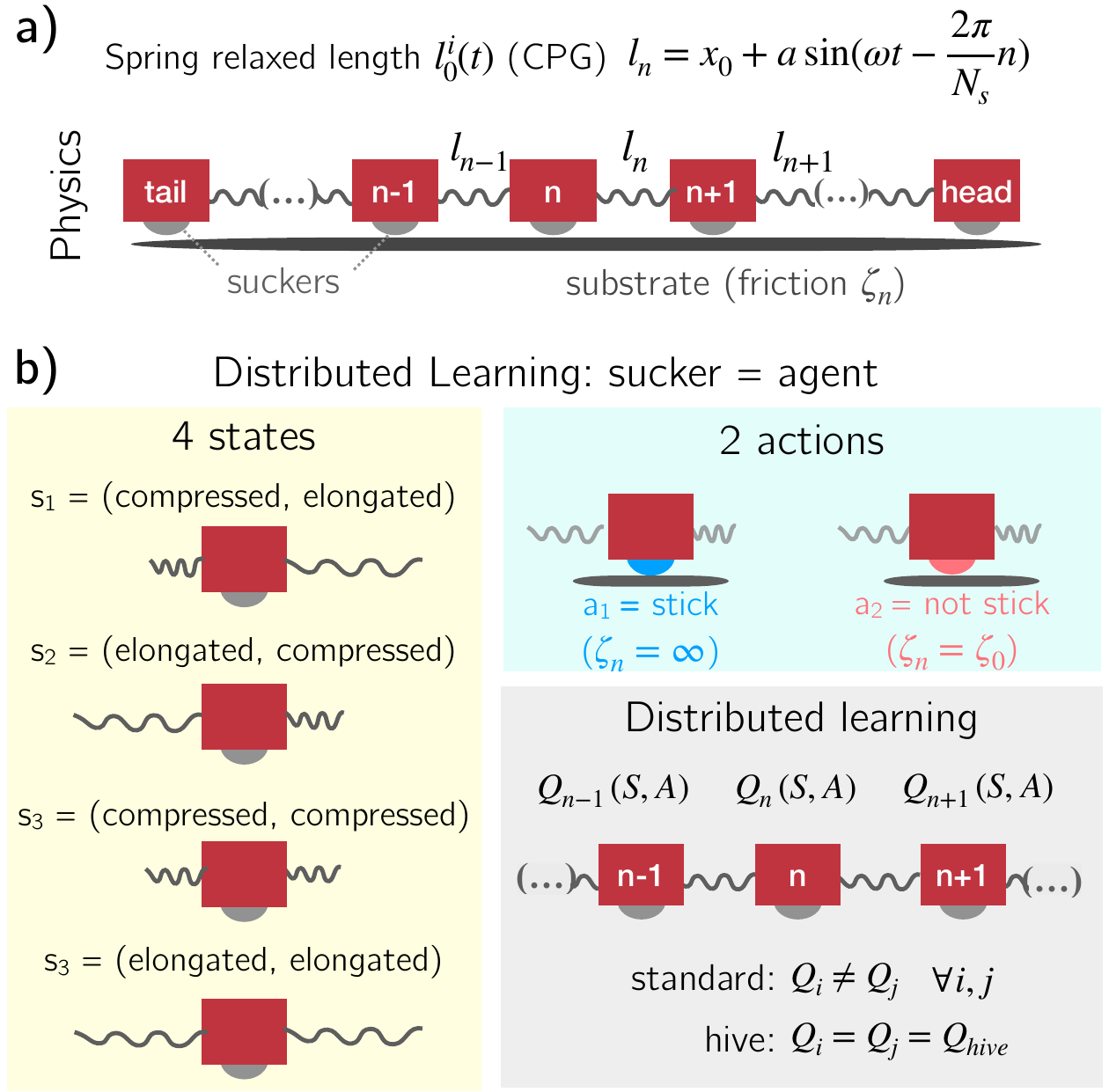}
    \includegraphics[width=\linewidth]{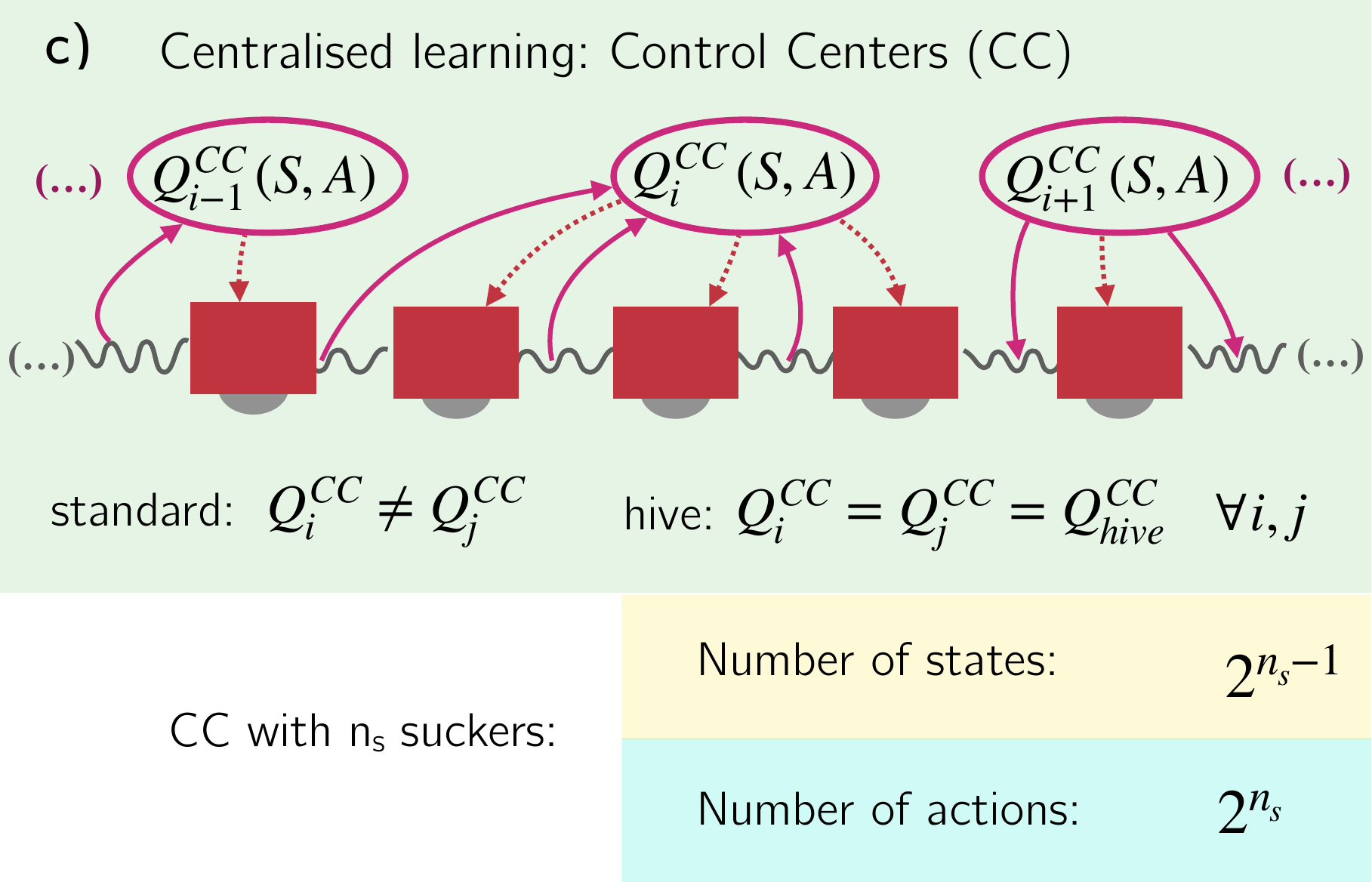}
\caption{a) Sketch of the 1D crawler model. b) Sketch of the distributed architecture, including the 4 states available to each agent (individual sucker), the 2 actions the sucker controls and the 4$\times$2 Q-matrix for each sucker. c) Centralized architecture: one or more Control Center (CC) control a subset of contiguous suckers. The states for each CC correspond to all combinations of compression/elongation states of the springs within the CC; the actions are all combinations of adhesion/no-adhesion of suckers within the CC. 
For multiagency (whether the individual agents are suckers or CCs), the \enquote{hive update} corresponds to forcing all agents to agree, i.e.~write and read the same $Q$ matrix. 
\label{fig:draft}
}
\end{figure}

\subsection{The 1D crawler model}


The model, illustrated in \cref{fig:draft}, is strongly inspired by the one proposed in \cite{Tanaka2012} and considers the crawler as a series of suckers (blocks) connected by springs along the body axis. The crawler is placed on a flat substrate and its dynamics is dictated by the springs and the interaction with the substrate:

{
\begin{equation}
\begin{split}
    m\ddot{x}_n(t) = -\zeta_n\dot{x}_n(t) + \kappa\left(x_{n+1}(t)-x_n(t)-l_n(t)\right)&\\
    -\kappa\left(x_n(t)-x_{n-1}(t)-l_{n-1}(t)\right)
    \end{split}
    \label{eq:springs}
\end{equation}
\noindent 
where $x_n$ is the position of the $n$th sucker, with $n=1,\dots, N_s$; $N_s$ is the number of suckers; 
$m$ is the mass of each sucker, $\zeta_n$ its friction coefficient 
and $\kappa$ the spring constant; $l_n$ is the rest length of the $n$-th spring and it oscillates in time as prescribed by the CPG as described below.
We assume friction is proportional to the sliding velocity of the suckers, modeling viscous friction with a wet substrate. Each sucker can increase its viscous coefficient $\zeta_n$ by adhering with the substrate. 
For simplicity, we assume that each friction coefficient can select only two values: a constant value $\zeta_0$ (no adhesion) or $\zeta_n\rightarrow \infty$ (the $nth$ sucker adheres). The choice of infinite adhesion is one of the ingredients distinguishing our model from \cite{Tanaka2012} where calculations are conducted under the assumption that individual suckers only slightly perturb their friction coefficient. 

For the tail, $n=0$ , and the head, $n=N_s$, the dynamics is driven by a single adjacent spring:
\begin{align}
    \label{eq:spring_head}
    m\ddot{x}_{0}(t) &= -\zeta_0\dot{x}_0(t) + \kappa\left(x_{1}(t)-x_0(t)-l_0(t)\right)\\
    m\ddot{x}_{N_s}(t) &=-\zeta_0\dot{x}_{N_s}(t) -\kappa\left(x_{N_s}(t)-x_{N_s-1}(t)-l_{N_s-1}(t)\right)
    \label{eq:spring_tail}
\end{align}

\subsection{Central pattern generator (CPG)}

We assume an elongation-contraction wave is generated by a CPG and propagates from the tail to the head of the crawler:
\begin{equation}
    l_n = x_0 + a \sin(\omega t - kn)
    \label{eq:CPG}
\end{equation}
\noindent where $l_n$ is the rest length of the $n$-th spring, defined as the one to the right of the $n$-th sucker; $k=2\pi/N$ with $N$ the periodicity, $\omega$ the angular velocity,  
$a$ a constant and uniform amplitude, and $x_0$ the natural length of the unperturbed spring. The unperturbed rest length of the springs, $x0$, fixes the spatial scaling of the model. The spatial periodicity $N$ in the CPG oscillation is in principle an arbitrary integer. For simplicity,  we fix $N=N_s$, so that the  wavelength $\lambda = x_0 N$ corresponds to the length of the crawler $L$. 
Arbitrary values of the periodicity are considered in Appendix B, where we show how $N$ alters the importance of each sucker when a crawling strategy is played.
Note that when all suckers are controlled by a single agent, it is possible to learn the contraction patterns for coordinated crawling without assuming a CPG, as shown in Ref.~\cite{Mishra2020}. We assume the CPG is unaffected by the suckers, relevant to real locomotion in the case where proprioceptive feedback has low gain~\cite{proprioception}. \\

\subsection{Scaling and parametrization}

The above equations can be non dimensionalized by rescaling all lengthscales with the unperturbed rest length $x_0$, $\bar{x}=x/x_0$, $\bar{l}=l/x_0$, and time with the frequency of the CPG, $\bar{t}=\omega t$ .
%
In the overdamped limit, inertia is negligible compared to the other terms, $m\omega^2/\kappa\ll 1$ 
and the dynamics reduces to kinematics:
\begin{equation}
    \begin{cases}
    \dot{\bar{x}}_n(t) = B\left(\bar{x}_{n+1}(t)+\bar{x}_{n-1}(t)-2\bar{x}_n(t)-\bar{l}_n(t)+\bar{l}_{n-1}(t)\right)\\
\dot{\bar{x}}_0(t)  = B\left(\bar{x}_{1}(t)-\bar{x}_0(t)-\bar{l}_0(t)\right)\\
\dot{\bar{x}}_{N_s}(t) = B\left(\bar{x}_{N_s-1}(t)-\bar{x}_{N_s}(t)+\bar{l}_{N_s-1}(t)\right)\\
\bar{l}_n = 1 + \bar{a} \sin( \bar{t} - 2\pi n/N)
    \end{cases}
\label{eq:model}    
\end{equation}
\noindent where the nondimensional group $B_n = \kappa/(\zeta_n\omega)$ is finite for suckers that do not adhere ($B_n=B_0$) and it vanishes for suckers that adhere ($B_n=0$ i.e.~when the $n$-th sucker adheres its velocity drops instantaneously to zero, $\dot{\bar{x}}_n =0$).
The problem is thus dictated by 4 nondimensional parameters:
\begin{equation}
    \begin{cases}
        B_0 &= \kappa/(\zeta_0\omega) = 10\\
        N   &= N_s \text{ except for \cref{fig:appendix_periodicity}}\\
        N_s &=L/x_0 \text{ varying}\\
        a/x_0 &= 0.25
    \end{cases}
    \label{eq:nondimensional_parameters}
\end{equation}
\noindent 

The model equations~\eqref{eq:model} with the non dimensional parameters~\eqref{eq:nondimensional_parameters}, the initial condition $\dot{x}_n=0$ and $x_n=nx_0$ and a specific scheduling for the control $\zeta_n$ (or its nondimensional form $B_n$) fully prescribe the dynamics. 
}
As detailed in \cite{Tanaka2012}, without coordinated adhesion of each sucker/block, the crawler just oscillates along its center of mass and the average velocity is zero. In contrast, when $\zeta_n$ is controlled appropriately, the crawler undergoes net translocation. In the following, we will compare the net translocation velocity $v$ under different strategies for the control of $\zeta_n$. 
To compare crawlers of different lengths, we will express velocities in units of the typical velocity $x_0\omega$: $\bar{v} = v/(x_0\omega)$. 

\subsubsection{Expected qualitative behavior from the idealized continuous infinite crawler}
\label{sec:analytical}
In Ref.~\cite{Tanaka2012} the authors find an analytical solution {for the velocity of the center of mass of a continuous ($N_s\rightarrow\infty$, $x_0\rightarrow 0$ and $Nx_0=\lambda$) one dimensional infinitely long crawler, assuming individual blocks can only perturb slightly the friction coefficient (rather than fully adhere like in our case)}. For the sake of clarity we summarize here their results. { The $n$-th friction coefficient is perturbed according to:
\begin{equation}
\label{eq:anchoring_pulse}
    \zeta_n(t) = \zeta_0(1+\epsilon p(s,t))
\end{equation}
where $s=nx_0$ is the continuous position along the crawler; $\epsilon\ll 1$ is a small parameter and only a single sucker can increase its viscous friction at a time, i.e.~$p(s,t)$ is only non-zero for a single sucker at a time. In Ref.~\cite{Tanaka2012} the authors show that to ensure fastest translocation, $p(s,t)$ should follow a wave traveling with the CPG eq.~\eqref{eq:CPG}, with a phase shift of $3\pi/2$. Under these optimal conditions, the center of mass translocates with a velocity:
\begin{equation}
\label{eq:vmax}
v\propto \frac{\tilde{\omega}}{\sqrt{1+\tilde{\omega}^4}}
\end{equation}
\noindent 
where $\tilde{\omega}=\omega\zeta_0N^2/(4\pi^2\kappa)$. Note that this equation has a maximum at $\tilde{\omega}=1$, which occurs when 
$$N_{\text{max}}=2\pi\sqrt{\kappa/(\omega\zeta_0)}.$$ 
\noindent This maximum is achieved when the wavelength of the CPG, $\lambda = Nx_0$, matches a diffusion lengthscale $\lambda_D=\sqrt{D/\omega}$ where $D=k x_0^2/\zeta_0$ can be derived analytically as the continuous model reduces to an effective diffusive equation. While our finite crawler cannot be treated analytically as in Ref.~\cite{Tanaka2012}, the diffusive lengthscale is still dimensionally correct, and we expect that qualitatively the crawler finds a maximum speed when $N_{\text{max}}\sim\sqrt{\kappa/(\omega\zeta_0)}$. 
We do indeed observe that the crawler translocates with a speed that depends on the number of suckers, with a maximum at an intermediate location with about $N=12$ suckers (Fig.~\ref{fig:results}). The location of the maximum varies with the parameters as prescribed by the scaling above (data not shown), suggesting that the physics of this optimal condition is robust to details of the model. In Appendix~A we discuss and compare extensively our model and simulator to Ref.~\cite{Tanaka2012}.\\
Note that living crawlers may evolve to match this constraint by e.g.~developing specific secretions that will modify their friction coefficient $\zeta_0$, or the frequency $\omega$ or wavelength $\lambda$ of their CPG, or the stiffness of their muscles $\kappa$. In the following, we do not focus on the mechanical parameters that define the crawler, but rather on how to learn the controls that enable it to move. 
}

\subsection{Learning architectures}
\begin{table*}
\begin{ruledtabular}
\begin{tabular}{lccc}
Architecture & $\epsilon$ & $\alpha$ & minimum training steps $\times$ episode \\
\colrule
Distributed (standard) & $\lesssim10^{-2}$ & $\sim10^{-3}$ &  $\sim50$k steps\footnote{difficult to converge and dependent on the number of suckers (agents)}\\
Distributed  (hive) & $\sim10^{-2}$ & $\sim10^{-3}$ & $\sim18k$ step\\
Centralized (small\footnote{5 or 6 suckers per control center.}CC) & $\sim10^{-2}$ &$\sim10^{-2}$ & $\lesssim100k$ steps, about half for hive\\
Centralized (large\footnote{10 or 15 suckers per control center.} CC) & $\sim10^{-2}$ &$\gtrsim10^{-1}$ & from $\gtrsim100$k to $\sim1000$k steps, about half for hive\\
Fully centralized (single CC) & $\sim10^{-1}$ & robust up to $\gtrsim 10^{-1}$& from $\gtrsim100$k to $\sim1000$k  steps \\
\end{tabular}
\end{ruledtabular}
\caption{ Approximate value of the exploration parameters used for the various architectures trained, including the number of integration steps per episode needed to obtain convergence (consider also \cref{fig:protocol}).
The precise value is case specific and was tuned manually by considering the trade-off between satisfactory levels of exploration and convergence.
\label{tab:explorationParameters}
}
\end{table*}

In the Reinforcement Learning (RL) paradigm, the learning problem is framed in terms of Environment and Agent. The agent, i.e.~the learner, interacts continuously with the environment selecting at each time step an \emph{action} (stick or not stick to the substrate) to which the environment responds with a numerical \emph{reward} and a representation of its \emph{state} \cite{Sutton2020}.
When treating the 1D crawler model as an RL problem, since the task is to achieve unidirectional and fast motion, the reward at each time step is given by the instantaneous center of mass velocity, $R_t = v_{CM}(t)$ when the velocity is positive, and $R_t=-1$ if $v_{CM}(t)<0$. After several trials, we found this choice to be effective to promote policies maximizing the crawling speed. 
In order to keep the problem as simple as possible and minimize the requirements on  proprioception, we use binary states  associated to the spring's tension: 1 = elongated; 0 = compressed. As previously {mentioned}, actions are also binary and related to each sucker being adhering or not: 
 $\zeta_n = \infty$ or $\zeta_n = \zeta_0$, respectively.
{Note that the instantaneous reward $R_t$ is a random variable, as the net speed of the center of mass is the result of the coupled spring dynamics which is only partially sensed and controlled by the crawler. We thus seek to maximize the cumulative translocation over many steps in the future, as usual in RL. To do this, we introduce the long term reward $G_t$, defined as $G_t = \sum_{k=0}^\infty \gamma^k R_{t+k+1}$ where $R_{t+k+1}$ is the instantaneous reward $k+1$ steps in the future with respect to $t$. We use the standard formulation with discount rate $\gamma<1$, which is a parameter discounting rewards further in the future ($\gamma=0$ truncates the sum to its first term i.e.~$G_t=R_{t+1}$ is the instantaneous reward; whereas $\gamma=1$ weighs each contribution equally for all times). We use $\gamma = 0.99$, corresponding to truncating the sum to an effective horizon of $100$ steps.}
For continuing tasks, another possible formulation of the reward is that of \emph{average reward} \cite{Sutton2020,Mahadevan1996}, which does not require discount and where the reward becomes also an estimate. 
Preliminary results however do not indicate any visible benefits of this formulation in our problem.

Learning {seeks to find a \emph{policy} of actions, $\pi(a|s)$, i.e.~a rule that prescribes how to choose actions $a$ at each given state $s$ that maximizes the expected future reward. Optimal policies} can be searched via several algorithms. We here use standard tabular Q-learning, where a \emph{quality matrix}, or Q-matrix, is learned with an iterative process. The Q-matrix {is the expected cumulative reward conditioned to being in state $s$ and choosing action $a$, i.e.~$Q(s,a)=E(G_t|S_t=s, A_t=a)$, while continuing with policy $\pi$. $Q$ is estimated iteratively and each estimate} is used to build a 
policy consisting in selecting at each state the action that maximizes the empirical return {with highest probability $(1-\epsilon_t)$ and choosing a random action with a small probability $\epsilon_t/|\cal{A}|$. Thus}  
for a given agent its Q-matrix follows the update rule
\begin{equation}
    \label{eq:Qupdate}
    \begin{split}
    Q_{t+1}(S,A) = &Q_t(S_t,A_t) +\\
    &\alpha_t\left [R_{t+1} + \gamma\max_a Q_t(S_{t+1},a) - Q_t(S_t,A_t)\right]\\
    \pi_{t+1}(a|s)=&\begin{cases}
    \arg\max_a Q_{t+1}(a,s) \,\, \text{w.p.}\,\, 1-\epsilon_t\\
    \text{any}\,\,\, a \in \cal{A} \,\,\text{w.p.}\,\, \epsilon_t/|\cal{A}|
    \end{cases}
    \end{split}
\end{equation}
{where $R_t$, $S_t$, $A_t$ are the reward, state and action of the agent at integration step $t$. The parameters $\alpha_t$ (learning rate) and $\epsilon_t$ (exploration) vary during learning in a process known as \emph{scheduling}, discussed in the next section.}
Therefore, the dimensionality of the Q-matrix is given by $|\mathcal{S}|\times|\mathcal{A}|$, with $\mathcal{S}$ and $\mathcal{A}$ the ensembles of states and action, respectively.
As illustrated in \cref{fig:draft}, agency can be defined by the learning architecture in several ways. Each architecture carries a specific definition of states and actions which dictates the dimensionality of the Q-matrix, hence the computational cost. 
{In particular, states are  binary words describing all possible combinations of elongation and compression of all $N_s-1$ springs contained within an agent. Spring $k$ is compressed when it is shorter than its rest length, i.e.~$\bar{x}_{k+1}-\bar{x}_k<\bar{l}_k$, where $\bar{x}_k$ and $\bar{x}_{k+1}$ and $\bar{l}_k$ are defined by eqs.~\eqref{eq:model}}.
Finally, we assume that each agent can act on temporal scales comparable to the CPG oscillation frequency, so that responses are instantaneous. 
To investigate the role of centralization \emph{vs} distribution, we explore two families of learning architectures, which we illustrate next.

\subsubsection{Distributed control}
Each sucker is considered an independent agent (panel b of \cref{fig:draft}). Each agent, excluding tail and head, has 4 states which are all the combinations of the spring states of the neighboring springs and 2 actions.
Tail and head of the crawler, have only 2 states since for them one spring is missing.
Therefore, each intermediate agent-sucker has a 4x2 dimensional Q-matrix, and tail and head a 2X2 Q-matrix.

\subsubsection{Centralized control}
We introduce centralization by defining an agent as an assembly of several suckers and springs which we call Control Center (CC), as illustrated by panel c of \cref{fig:draft}. 
The number of CCs controlling the crawler depends on how many suckers are included in each CC and can range from several CCs down to a single CC, which corresponds to the highest degree of centralization. 
Centralized architectures control several suckers and springs simultaneously, hence have access to a richer representation of the dynamical state of the crawler, and can realize many more policies than those realizable in the distributed setting.
The state of the CC is given by the compression/elongation state of each spring located between the first and last sucker within the CC; while the action of the CC consists of specifying the binary adhesion/no-adhesion instruction for each of the suckers within the CC. 
Since the CC has access to all combinations of compression/elongation and adhesion/no-adhesion states and actions, the state-action space scales exponentially as $|\mathcal{S}| = 2^{n_s-1}$ and $|\mathcal{A}| = 2^{n_s}$, with $n_s$ the number of suckers controlled by the given CC (for $n_s$ suckers under a CC there are $n_s-1$ connected springs). Note that to facilitate comparisons and have the same dimensionality of the Q-matrix for each CC 
we chose to ignore the spring between two adjacent CCs.  
{ This is the only consistent choice for hive updates, as if the spring at the interface of two control centres was included in the state, the information would be interpreted differently by the two adjacent control centres. To make a fair comparison, we keep the same choice even when each control centre learns its own policy (non hive). We do not expect considerable differences because the control centres already perform as well as the single agent, to within the error bar.}

\subsubsection{The hive assumption}
Whenever multiple agents control crawling, a simplifying assumption may be adopted imposing that all agents learn the same policy. This is the so called \emph{hive} assumption or Population-Based Training \cite{Terry2023,Plaat2022,Young2020}, where a population of agents is used to find a shared solution. 
In all our multiagent learning schemes --including distributed and centralized with multiple CCs-- we compare standard learning to hive learning, as illustrated in \cref{fig:draft}.  
In practice, each agent writes and reads the same Q-matrix. The distributed hive Q-matrix becomes a 8X2 tensor, where in addition to the 4 states associated to the internal sucker there are 2 extra {elements of the matrix} for the tail, (null,elongated) and (null, compressed), and for the head, (elongated,null) and (compressed,null)\footnote{Note that in this case the problem is effectively hive only for the internal suckers, whilst tail and head remain genuinely independent (since their states are unique) as in standard multiagent RL.}.
For the hive CCs there is no need to modify the Q-matrix between hive and standard learning, since all CCs have Q-matrices that are identical in shape, as discussed above.
Clearly, the hive assumption speeds up training as it allows to parallelize the learning process. However, it also affects significantly the learned policies, as all agents are forced to learn the same policy, despite being placed at different locations relative to the propagating wave. 
In other words, the hive update imposes a consensus mechanism and all agent must  \enquote{agree} on the same policy. 
In fact in the hive policy we find that different agents will learn distinct policies, and the hive update reduces the number of policies that are learned, as discussed in the Results section.

\subsection{Learning protocol}
\begin{figure}[h!]
        \includegraphics[width = \linewidth]{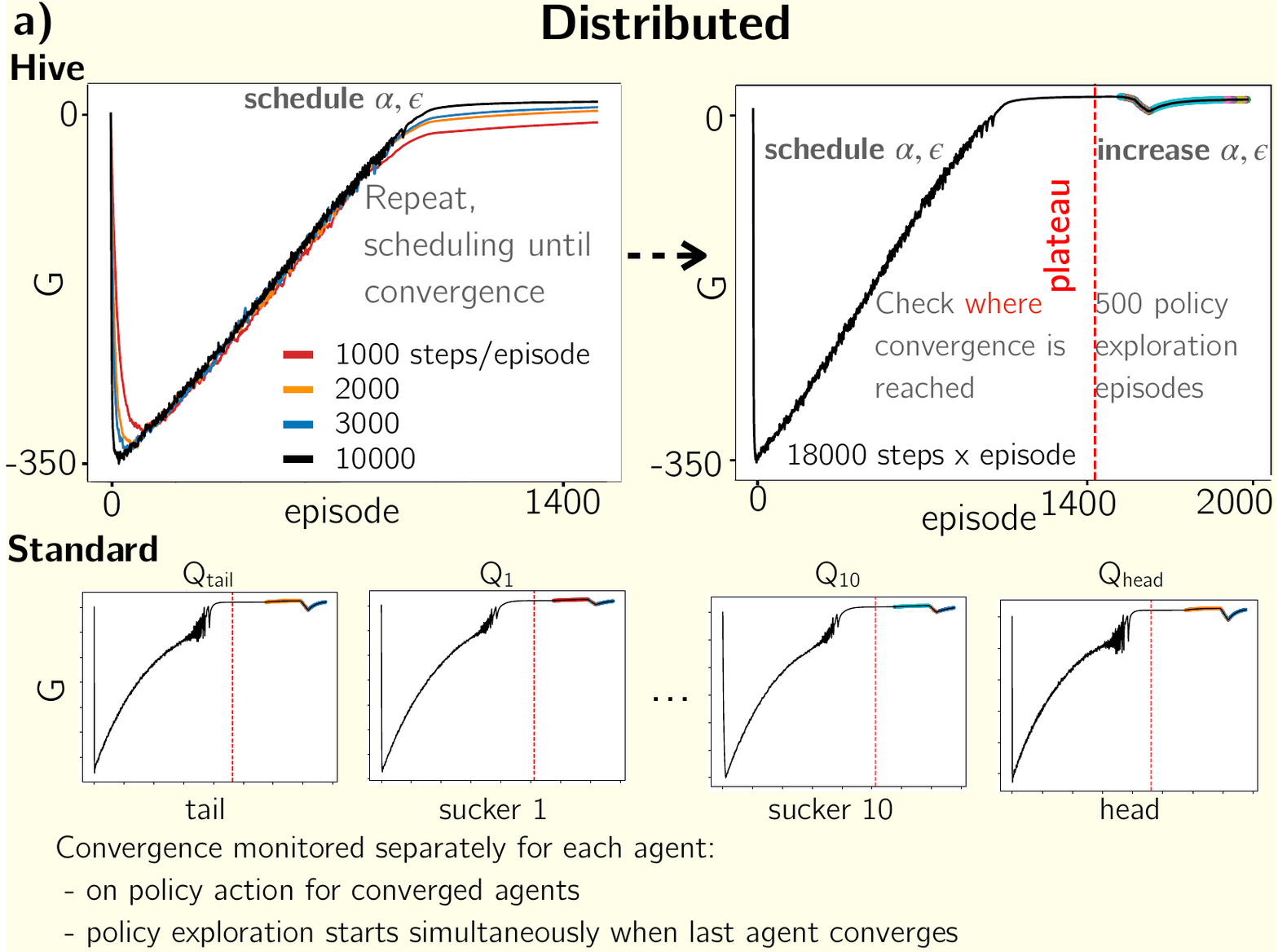}\\
        \includegraphics[width = \linewidth]{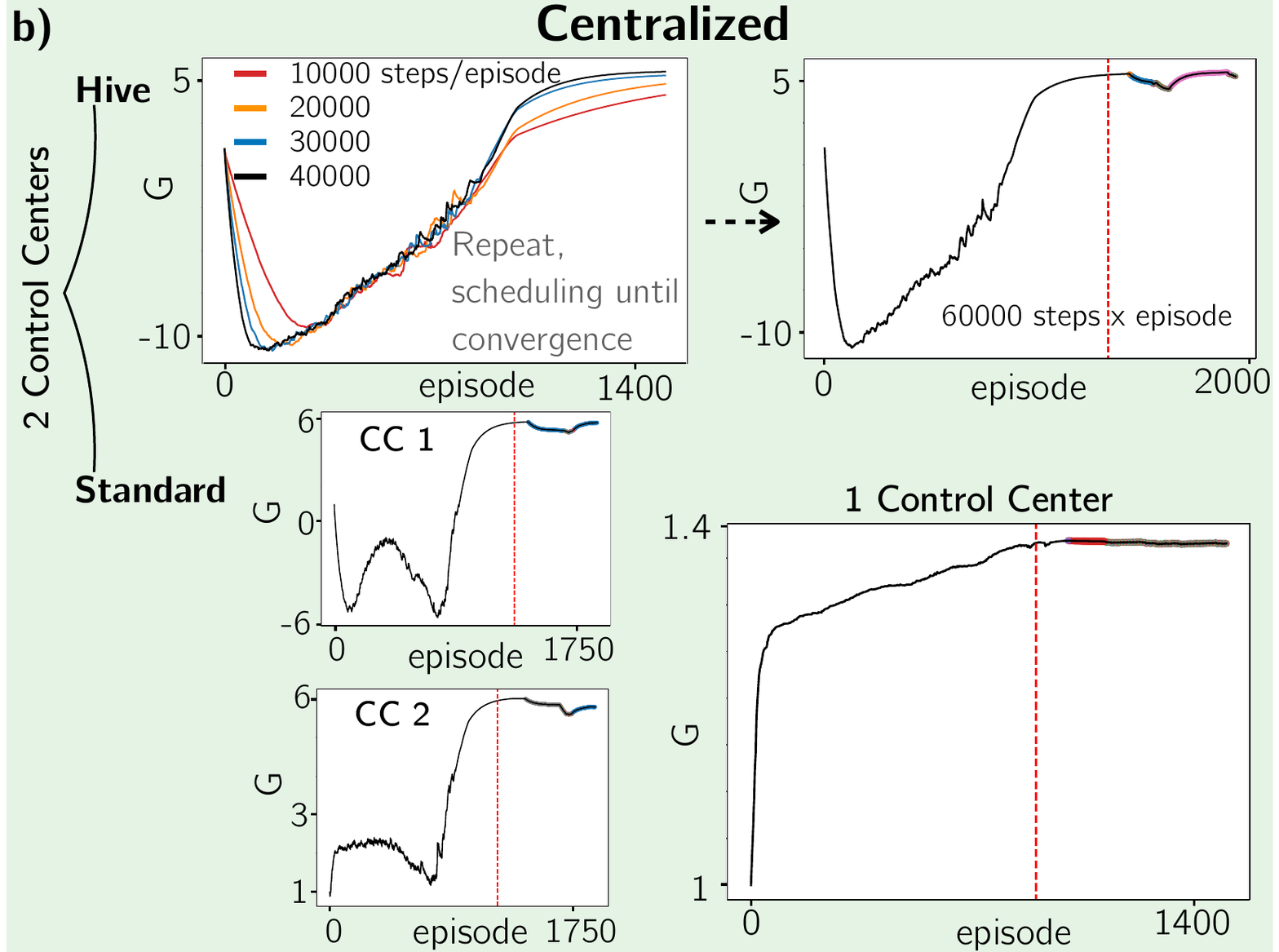}
    \caption{\label{fig:protocol}
    Illustration of the training protocol on a 12 suckers crawler for various learning architectures. (a) Average return {$G=1/|\mathcal{S}|\sum_s\max_a\left(Q(s,a)\right)$ as a function of learning episode for distributed learning architectures with the same Q matrix for all agents (hive, top) or one Q matrix per agent (standard, bottom). Top left: $G$ initially increases and then slows down; the process is repeated increasing the number of steps per episode (color lines) until learning reaches a flat plateau (black). Right: at the plateau, identified automatically (red dashed line),  $\epsilon$ and $\alpha$ are increased again within an exploration phase that stores a set of 250 policies (each color represents a different policy). (b) Same as (a), for centralized architectures with 2 control centers hive (top) and standard (bottom) and for a single control center (bottom, right).  See \cref{tab:explorationParameters} for details on the values of the parameters.} 
    }
\end{figure}
RL theory and algorithms are extensively covered in specialized books and articles (e.g.~\cite{Sutton2020}). 
As briefly summarized before, in tabular Q-learning during training we iteratively update the Q-matrix using \cref{eq:Qupdate}.
To keep a certain degree of exploration during training we adopt the standard procedure of following an $\epsilon$-\emph{greedy} policy, that is, with a finite probability $\epsilon$ a random action is chosen instead of the one prescribed by the current policy (i.e.~the action maximizing the current estimate of the value).
Duriring training, both the exploration $\epsilon$ and learning rate $\alpha$ in \cref{eq:Qupdate} are scheduled to decrease linearly, to ensure convergence to a final estimate of Q. Note that convergence to the optimal Q matrix is only guaranteed asymptotically, hence different trainings will lead to different sub-optimal policies, which we explore systematically in the following.
In this work the scheduling process starts at a value of $\epsilon_{\max} = 0.9$ and $\alpha_{\max} = 0.1$ and is scheduled down to $\epsilon_{\min} = 0.001$ and $\alpha_{\min} = 0.001$.
Because here we seek fast unidirectional motion, the task is \emph{continuing} and there is no termination.
With the term \emph{episode} {we intend a series of updates characterized by the same value of the parameters} 
$\alpha$ and $\epsilon$. Note that this definition is different from the usual definition used in episodic tasks, where the end of the episode corresponds to the agent reaching a specific terminal state. 

Note that although the equations of motion are deterministic, our learning problem is stochastic. There are two sources of stochasticity: i.~the  coarse grained state space and ii.~the definition of agents. Both i.~and ii.~imply that the reward signal is a stochastic variable. 
Indeed, the speed of the crawler depends on the full dynamical state of the system, defined by the (continuous) speed, position and adhesion of all suckers and on the relaxed length of all springs. However, as per i.,~our crawler does not observe the full dynamical state of the system, but rather its coarse projection onto discrete binary states of compression vs elongation of all the springs. Thus given the actions of all suckers, the center of mass is not fully determined yielding a stochastic reward $R(t)$. 
Moreover, in all multiagent architectures each agent is unaware of other agents' actions, as per ii., hence once again the actions of any individual agent do not individually determine the speed of the center of mass (i.e.~the reward). 
This latter source of noise in the reward is bypassed by the fully centralized architecture with a single control center (1CC).

To monitor progress during learning we consider the average value $G$, with $G=1/|\mathcal{S}|\sum_s\max_a\left(Q(s,a)\right)$. 
{Clearly, when $G$ increases, the $Q$ matrix is being updated and learning is in progress, whereas when $G$ reaches a plateau, the learning process has come to an end.} As illustrated by \cref{fig:protocol}, we set an automatic protocol consisting in repeating the learning procedure until a plateau in $G$ is detected: when the plateau is not encountered  after 1500 episodes, we repeat the process changing the scheduling by increasing the number of steps per episode. When several Q-matrices are learned simultaneously (for instance, in standard distributed control) the convergence is monitored separately for each agent.
Due to the stochastic nature of the problem, we noticed that repeating the learning procedure yields different policies. 
Therefore, we design a protocol to account for the stochasticity of this learning process. To efficiently store several sub-optimal policies, without repeating the entire training procedure, we introduced the following strategy: when convergence is detected, the learning parameters are raised to a small and constant value and 500 extra episodes are run letting Q evolve accordingly. At each episode the corresponding policy is stored.
As reported in \cref{tab:explorationParameters}, the value of $\alpha$ and $\epsilon$ for this \enquote{policy exploration} phase are set differently according to the learning architecture considered with the purpose of exploring (sub) optimal policies without disrupting significantly the plateau that was reached by $G$.
The value of $\alpha$ and $\epsilon$ that is needed to remain in the plateau reflects how robust the optimal policy is to perturbations, and appears correlated to the degree of stochasticity previously discussed. Centralized architectures appear to need stronger exploration of the plateau to {learn distinct} sub-optimal policies. We find a fairly stable plateau region in the exploration phase of the fully centralized 1 CC architecture even for large values of $\alpha$ and $\epsilon$. This is due to the high degeneracy of this architecture, where many different policies perform similarly, as discussed later (see \cref{fig:pol_distr}). 
Finally, we find that the hive update introduces some degree of stabilization of the learning process since we observed the need of less integration steps (as expected) and a smaller sensitivity to scheduling and exploration parameters. The source code for all different architectures is available at~\cite{githubSourceCode2025}.
{Next we discuss results obtained by repeating the protocol described above for several centralizes and distributed architectures and a number of suckers that ranges from $N_s=5$ to $N_s=30$. We reserve a particularly in-depth analysis to the case of $N_s=12$.}

\section{Results}
\begin{figure*}
    \centering
    \includegraphics[width =0.85\linewidth]{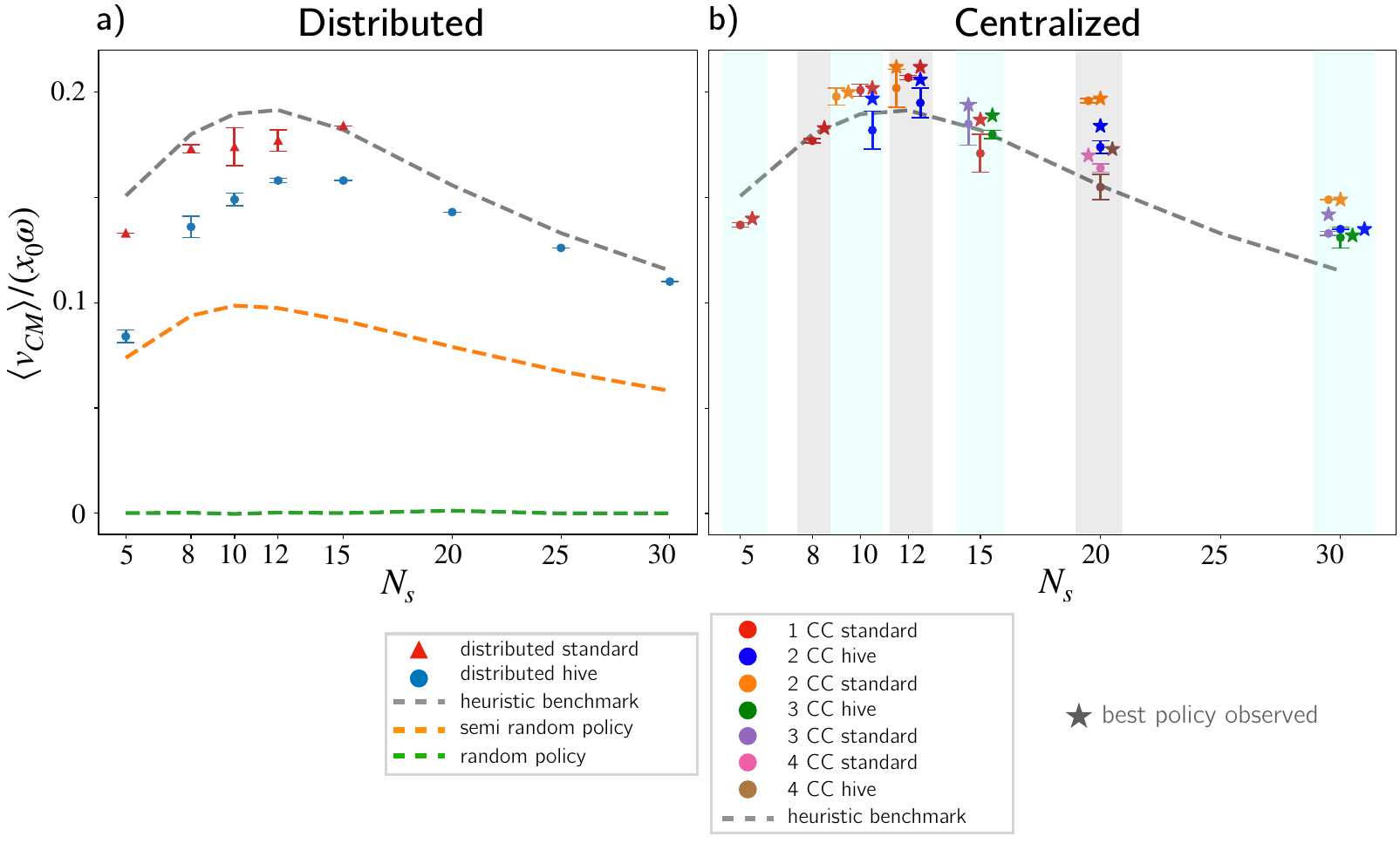}
    \caption{Average velocity of the crawler center of mass \emph{vs} number of suckers, for different learning architectures. Markers and errorbars correspond to average and standard deviation of performance over the 250 best learned policies for distributed architecture (left) and centralized architectures (right). Colored bands in panel b) group results obtained for the same value of $N_s$ (which have been horizontally scattered for clarity, so symbols do not overlap).
    \label{fig:results}}
\end{figure*}
Using the learning protocol described above, we store 500 optimized policies for each architecture and quantify performance as the average speed of the center of mass of the crawler for the best 250 policies. 
We obtained velocities by playing each policy over a trajectory of 20000 steps, or about 300 full cycles of the CPG travelling wave across the full crawler. In computing the  average we accounted for the multiplicity of each policy, so that policies which are found several times will be more represented as well as quasi-degenerate policies realizing similar crawling velocities.

We find that performance for all architectures depends on the number of suckers $N_s$ (\cref{fig:results}), qualitatively consistent with the infinite continuous crawler case previously discussed. { With our choice of parameter,} performance peaks for $N_s\approx 12$ and then exhibits a long-range decay beyond the peak (as discussed in Methods, the number of suckers is identical to the spatial periodicity of the CPG wave, $N$). 
To investigate the limits and properties of multiagency and centralization in cooperative/coordinated actions we will focus largely on the peak. 
{The value of the parameters we chose correspond to the peak occurring near $N_s=12$, which is computationally treatable for all learning architectures from the most distributed to the most centralized.}
Whether and how natural crawlers orchestrate the central pattern generation and the suckers to operate near the peak is a fascinating question for further research. 

We find that the tail and head of our finite crawler play a key role, thus distinguishing our problem from the infinite crawler for which an asymptotic solution is available. However, the analytical optimal policy is a useful heuristic for the internal suckers as discussed below. 
We thus construct a reference heuristic policy prescribing that \emph{(i)} internal suckers follow the analytical optimal policy~\cref{eq:anchoring_pulse} 
from \cite{Tanaka2012} (see [Supplementary Video 1]), and  
\emph{(ii)} tail and head anchor in response to negative forces, i.e.~the head adheres when it is dragged backward by the elongated spring to its left and the tail adheres when it is pushed backward by the compressed spring to its right. It is important to observe that to implement the heuristic strategy, the simulator prescribes adhesion according to a phase shift, thus measuring time and space. In contrast, because we use binary compressed/elongation spring states, our agents (the suckers, or the CCs) have no access to the CPG's wave phase. 
Thus, our simple crawlers cannot implement the heuristic policy, due to their limited states and controls. Intriguingly, our poor state representation can be compensated by a more centralized control which recovers a smooth traveling wave by controlling multiple suckers at a time, as we will see in the following.  

\subsection{Distributed control}
Let us first focus on distributed forms of control, where each sucker is an agent (\cref{fig:results}, left). 
When suckers play random actions, the center of mass does not move on average, as expected (random policy, green dashed line). If tail and head adhere in response to negative forces, even if the internal suckers perform a random action, the center of mass does move, showing the importance of the crawler's tail and head (semi-random policy, orange dashed line). If tail and head adhere in response to negative forces and the internal suckers adhere according to the analytical results described above, performance further increases (heuristics, gray dashed line). 
Although the analitically optimal policy of the internal suckers is obtained in the infinite continuous crawler limit, the clear improvement of the policy relative to the semi-random policy suggests that away from the boundary, the infinite crawler may be a useful approximation of our finite discrete crawler. { This is consistent with the fact that our  crawler translocates with a speed that varies with the number of suckers, qualitatively similar to the infinite crawler in Ref.~\cite{Tanaka2012}. We report this heuristics as a qualitative reference to guide the eye. Note that experiments on leech~\cite{leech} support the role of head and tail postulated here. }
When we let agents act according to a policy that has been learned through trial and error, performance is intermediate between the semi random policy and the heuristics ({\cref{fig:results}a}, blue circles and red triangles). We find that standard learning is harder than hive learning, but it achieves better performance and nearly matches the heuristics described above (e.g.~for 15~suckers). As a reminder, in hive learning all suckers learn the same policy, whereas in standard learning each sucker learns its own individual policy. 
\\
Almost invariably trainings with hive architectures learn the same policy $\pi_H$, regardless of $N_s$. 
According to $\pi_H$, internal suckers adhere in response to the $\leftarrow|\leftarrow$ state while the tail and head --which are connected to a single spring-- adhere in response to the states $\times|\leftarrow$ and $\leftarrow|\times$ respectively. 
The interpretation of the hive policy $\pi_H$ is simple: each sucker adheres in response to negative forces, i.e.~when all adjacent springs push it backward. 
For $N_s\ge 15$, all our trainings converge to the same policy $\pi_H$, with no suboptimal policies ever learned (hence the absence of the errorbar in~\cref{fig:results}a). 
The same behavior was found when training distributed hive policies for even larger crawlers with $N_s >30$ (results not shown)
further confirming that $\pi_H$ is the best consensus policy. Departures from this policy emerge only for small $N_s$ where the specific dynamics of few suckers might affect the global consensus.
For $N_s < 15$ trainings converge to a small number of policies; the best learned policy is $\pi_H$ for $N_s=5$ and $12$, and it is a slightly different policy $\hat\pi_H$ only for $N_s=8$ and $10$. $\pi_H$ and $\hat\pi_H$ are almost identical, except internal suckers also adhere in response to the $\rightarrow|\rightarrow$ state, and the tail never adheres. 
For small number of suckers, $\hat\pi_H$ is marginally faster than $\pi_H$ (e.g.~ for 10 suckers speed is $\approx0.017$ vs $\approx0.015$).
Thus collective hive learning is highly transferrable, in that the hive policy can be applied to crawlers with any number of suckers with no need for retraining, consistent with previous results obtained with genetic algorithms~\cite{Hartl2025}.
Adopting the hive policy results in a clear traveling wave of adhesion, with only one sucker adhering at a time (see animation of the hive policy in [Supplementary Video 2]). Note that, as already remarked above, the individual suckers do not have access to their phase shift, and as a result, this architecture cannot learn a policy that implements our heuristic benchmark. Indeed, this heuristics requires a finer proprioceptive signal, e.g.~inferring the phase from the precise intensity of the restoring forces, rather than its sign. 

Intriguingly, the standard learning, where each sucker acts according to its own policy, realizes a more complex dynamics with multiple suckers adhering simultaneously and no visible traveling wave of adhesion. In doing so it achieves a more complex group behavior and better performance.
Note that, although standard learning achieves better performances, we find that training is more computationally expensive and in our hands was limited to $N_s=15$: beyond this limit the training did not converge. This is likely due to the intrinsic level of stochasticity: all agents learn simultaneously, but rewards are coupled, as the motion of the {crawler} depends on the full dynamical state of the system. Thus good actions may be hard to learn as they may not be individually good enough to yield a positive reward.
Moreover, in all cases where standard learning was successful, we noticed an extreme sensitivity to the scheduling parameters, with convergence only using about three times the number of steps per episode relative to the hive solution. In contrast, the distributed hive policies were easily learned.

\begin{figure}
    \centering
    \begin{subfigure}[t]{\linewidth}
        \includegraphics[width = \linewidth]{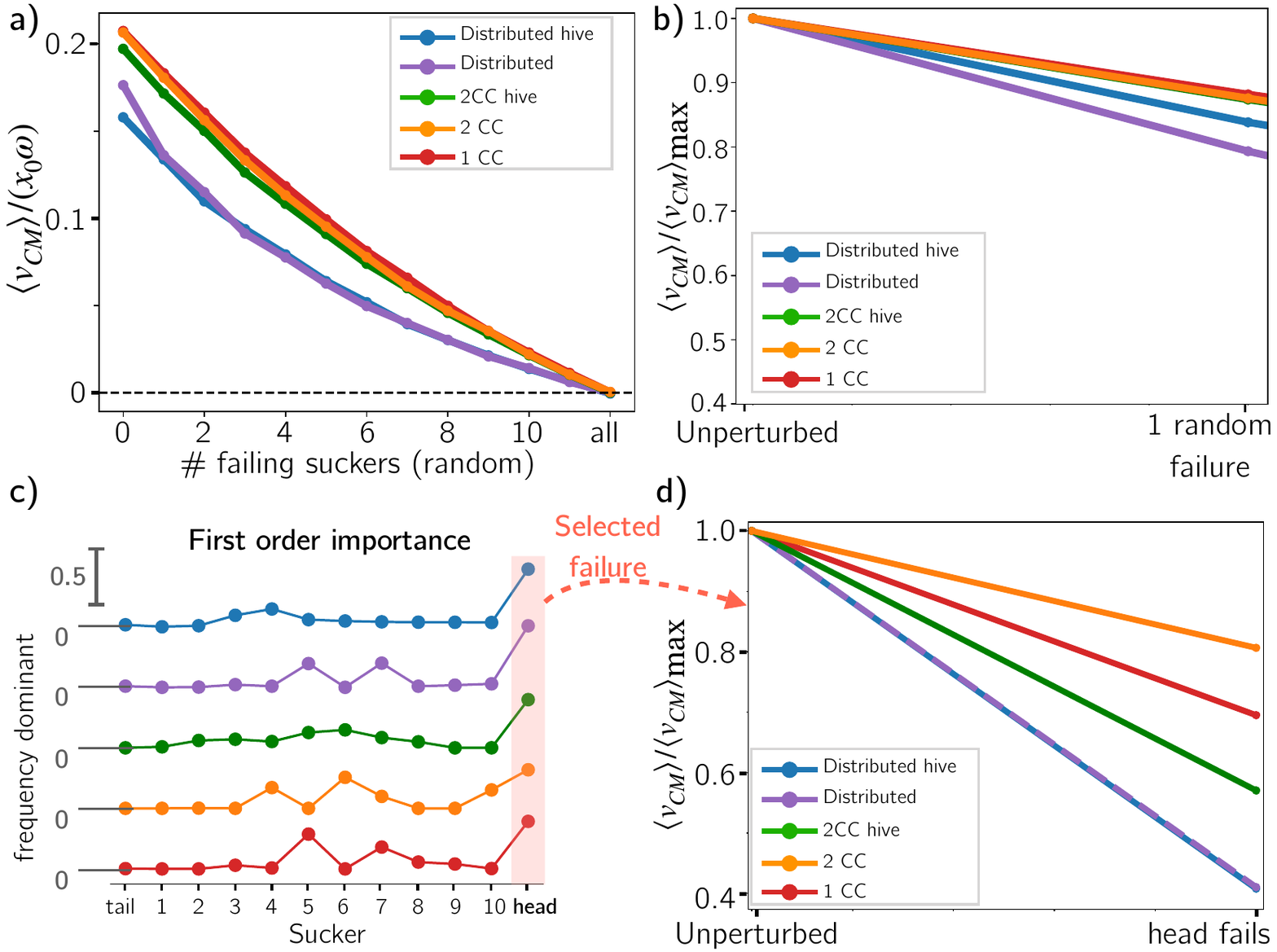}
    \end{subfigure}
    \begin{subfigure}[t]{\linewidth}
        \includegraphics[width = \linewidth]{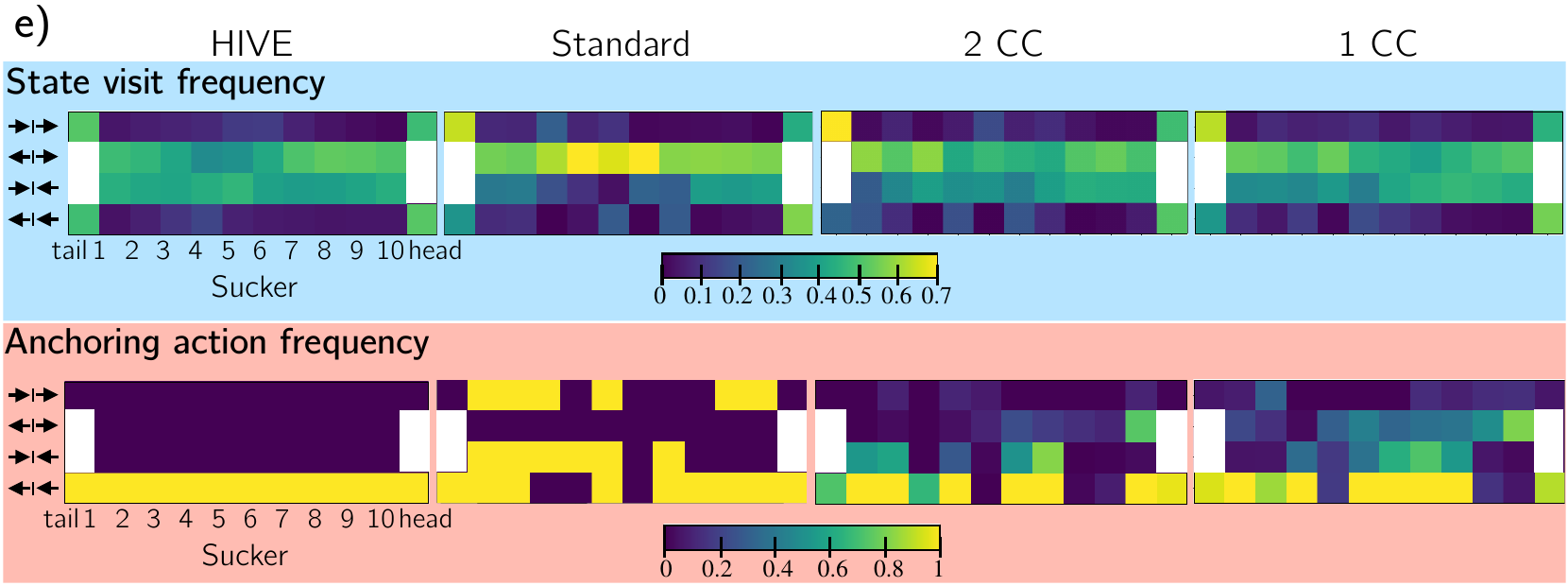}
    \end{subfigure}
    \caption{Robustness analysis of the best policy for 12 suckers crawlers with various learning architectures. a) Performance (velocity of the crawler) \emph{vs} number of failing suckers. 
    b) Performance of the optimal policy with 1 failing sucker relative to optimal performance with no failing suckers.  
    c) Frequency each sucker is the dominant contributor, { measured by the largest drop in performance when the sucker fails}: the head is the most important sucker.  
    d) Performance drop due to failure of the head. 
    e) Statistical behavior of a 12 sucker crawler for different distributed and centralized architectures. Top row: for each sucker (x axis), and each compression state of its adjacent springs (y axis), the color represents the frequency that state was visited while performing the optimal policy. 
    Bottom row: Fraction of time a sucker adheres, colorcoded from purple (always adhere) to yellow (never adhere) for each of the 12 suckers (x axis) and each of the compression states of their adjacent springs (y axis). 
    The blank entries for head and tail represent that these suckers only have access to one spring, hence two compression states (rather than four).  
    \label{fig:robustness}}
\end{figure}
\subsection{Centralized control}

Centralized architectures achieve even better performances, at the cost of increased computational burden (\cref{fig:results}, right).  
Indeed, the number of entries in each Q matrix grows exponentially with the number of suckers controlled by the corresponding control center (CC). Given that $n_s\leq N_s$ is the number of suckers controlled by each control center, there are $N_s/n_s$ CCS whose quality matrix has dimensions $|\mathcal{S}| \times |\mathcal{A}| = 2^{n_s-1}\times 2^{n_s}$.
Consistently, learning requires considerably longer training (\cref{tab:explorationParameters}) relative to distributed architectures, where training requires to optimize $N_s$ Q-matrices whose dimensions are independent from the crawler size $|\mathcal{S}| \times |\mathcal{A}| = 4\times 2$. 

At the same time, a higher degree of centralization achieves better performance. We find that a single control center dictating the actions of all $N_s$ suckers performs best for $N_s = 10$ and 12, suggesting that 
long range correlations across the whole crawler are useful, if well orchestrated. However, training of the single CC is extremely complex; it failed beyond $N_s=15$, because of the sheer size of the quality matrix, and for $N_s=15$ it under performs slightly relative to the 3 CCs. Interestingly, multiple CCs achieve almost the same performance as a single control center, and can be trained even with a much larger number of suckers. For example, the reduction in complexity for $N_s=12$ is quite impressive: the Q-matrix for a single CC has approximately $10^7$ entries, whereas it shrinks to about $2000$ entries with two control centers. 
As for distributed learning, hive learning where each CC is forced to learn the same policy is faster than standard learning, but performs slightly more poorly, although the margin is not as impressive as for the distributed control. 
Next, let us illustrate in more detail the differences between these architectures, focusing on the 12 sucker crawler. 
For illustrative purposes, the optimal policies for the various learning architectures, both distributed and centralized, are shown for a 12 suckers crawler in [Supplementary Video 3].

\subsection{Robustness}
We showed in the previous sections that moving from pure multiagency to some form of centralization improves performance. However, the benefits are not always sizable, e.g.~1CC improves $\sim17\%$ over distributed standard and $\sim30\%$ with respect to distributed hive, when considering 12 suckers. 
At the same time, computational cost grows dramatically and optimal policies become hard to learn.
Are there any other benefits to centralized control? \\
We next show that centralized architectures are considerably more robust to failure of individual suckers. 
In \cref{fig:robustness}, we focus on the best policy obtained for each architecture for the 12 suckers crawler and test its robustness to failure. To this end, we extract one or more random suckers at each integration step, and let them perform a random action (adhere or not with the same probability). All other suckers follow the optimal policy, and no extra learning is allowed. Centralized architectures are more robust to failure: their performance { drop more slowly} with the number of failing suckers \cref{fig:robustness}(a), 
with a single random failure eroding less than 10\% performance \emph{vs} about 20\% for distributed architectures, \cref{fig:robustness}(b). 
We then asked how robust these policies are to failure of the most crucial sucker.  
When we make a single sucker fail consistently  over time, we find that the head plays the most prominent role in all of our architectures, as its failure contributes the largest drop in performance \cref{fig:robustness}(c). 
In fact the head is always the dominant sucker when the CPG wavelength, $\lambda$, is equal or larger than the crawler length and not otherwise (appendix C). 
Failure of the head is particularly acute for distributed architectures (60\% drop in performance for fully distributed architectures) against much less dramatic drops in centralized architectures (the most robust being the 2CC with a $<20\%$ drop).\\
To further quantify the distinct policies we measure the frequency that each sucker experiences the four local compression/elongation states of its adjacent springs, as the policy is played over time. The four local states are represented with the symbols $\rightarrow|\rightarrow$; $\leftarrow|\rightarrow$; $\rightarrow|\leftarrow$ and $\leftarrow|\leftarrow$. The local states are visited with frequencies that vary across the different architectures (\cref{fig:robustness} (e) top row), but the pattern is relatively preserved. This suggests that the underlying physics constrains the spring dynamics considerably, and may partly overshadow differences in the control architecture. In contrast, the average action played by each sucker in its local compression/elongation state varies dramatically across different control architectures (\cref{fig:robustness} (e) bottom row). In distributed architectures (left panels), suckers play an action that depends only on their local compression/elongation state, hence colors are binary (adhere or not adhere). In contrast, in centralized architecture, a sucker acts in response to the global (albeit binary) state of compression/elongation, and not only in response { to} its local state. Intermediate colors in the right panels signify that the local compression/elongation state is not enough to determine whether the sucker will adhere or not. Because centralized policies are dictated by global compression/elongation states, they may be more tolerant to failures of individual suckers.

\begin{figure}[h!]
    \centering
    \includegraphics[width=\linewidth]{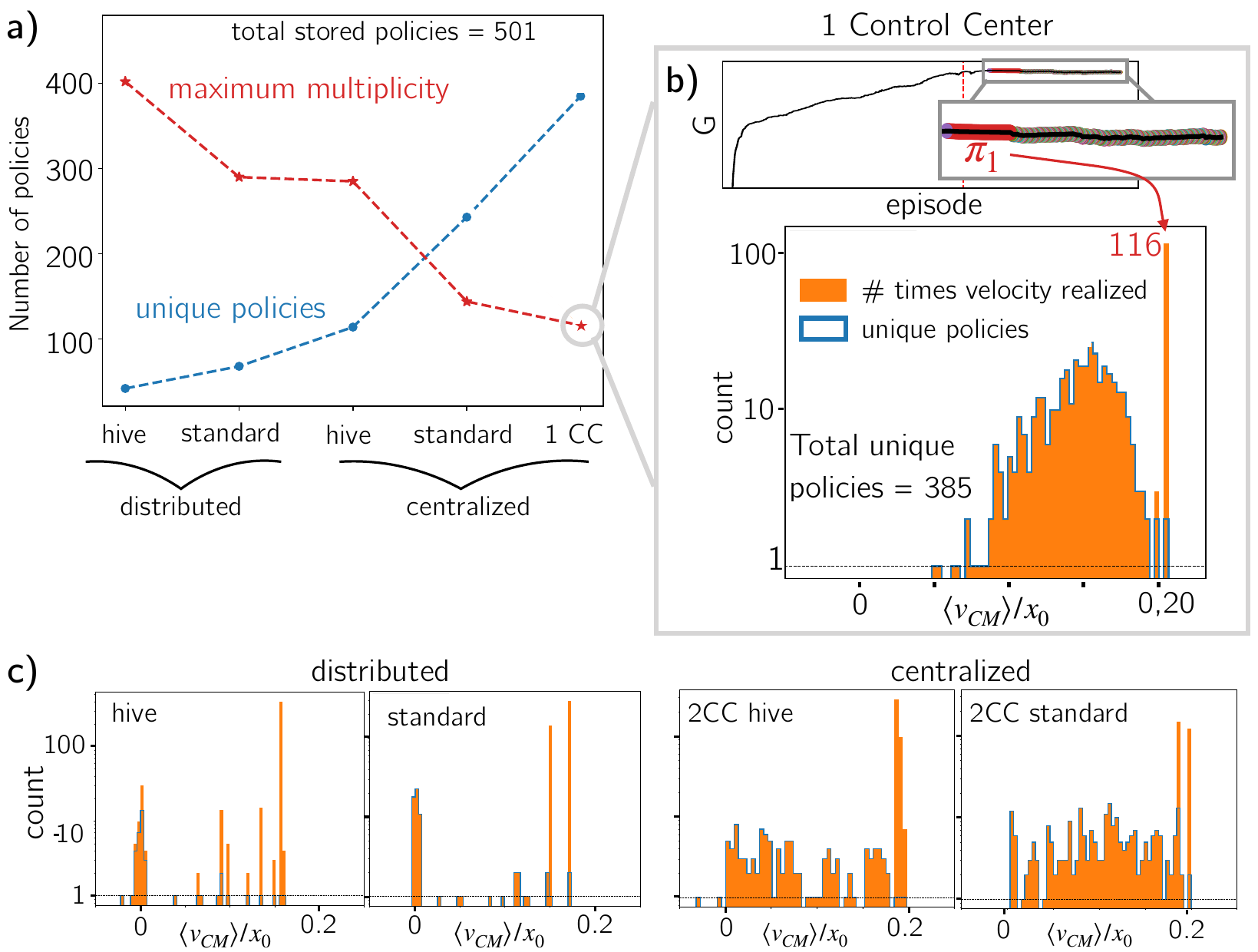}
    \caption{\label{fig:pol_distr}
    Distribution of speed for the policies learned by the 12 suckers crawler. 
    {a)}: number of distinct policies learned by the 12-sucker crawler with different architectures (blue); number of times the most popular policy was learned (red);  {b)}: detail of policies learned by the fully centralized agent; top: plateau in $G$, showing { the optimal policy, labeled as $\pi_1$, is selected at several iterations of the exploration phase of the learning protocol
    (same as bottom right panel in fig.~2b, color change in the plateau marks policy change); bottom:
    histogram of the velocities across all 250 policies. Orange: number of times a crawler velocity is realized by a learned policy . Blue: number of unique policies associated to the velocity (within the bin).
    Bins where the blue histogram is lower than the orange histogram show identical policies which are selected multiple times in the learning protocol.
    c): Same as b), for the remaining learning architectures.}
    }
\end{figure}
\subsection{Redundance}
To further elaborate on why centralized policies are more robust, we characterize redundancy.
Within a given learning architecture, are many sub-optimal policies learned and are they considerably worse than the best policy? 
Centralized learning converges several times to the optimal policy but also to several fair sub-optimal policies. For instance, the 1CC architecture converges 116 times out of 500 to the optimal policy $\pi_1$, but also converges to another 384 sub-optimal policies whose performance is only slightly poorer than the optimal performance (\cref{fig:pol_distr} top row). 
Conversely, distributed learning converges even more often to the best policy (large peaks in the histograms in \cref{fig:pol_distr} bottom row, corresponding to the best policy) and to fewer considerably worse sub-optimal policies (group of bins near velocity zero).
These results suggest that the higher degree of complexity of centralized control offers numerous ways to achieve fast crawling strategies making it harder for the algorithm to spot \emph{the} best one but at the same time more likely to learn \emph{a} good policy. 

\begin{figure}[h!]
    \centering
    \begin{subfigure}[t]{\linewidth}
        \includegraphics[width = \linewidth]{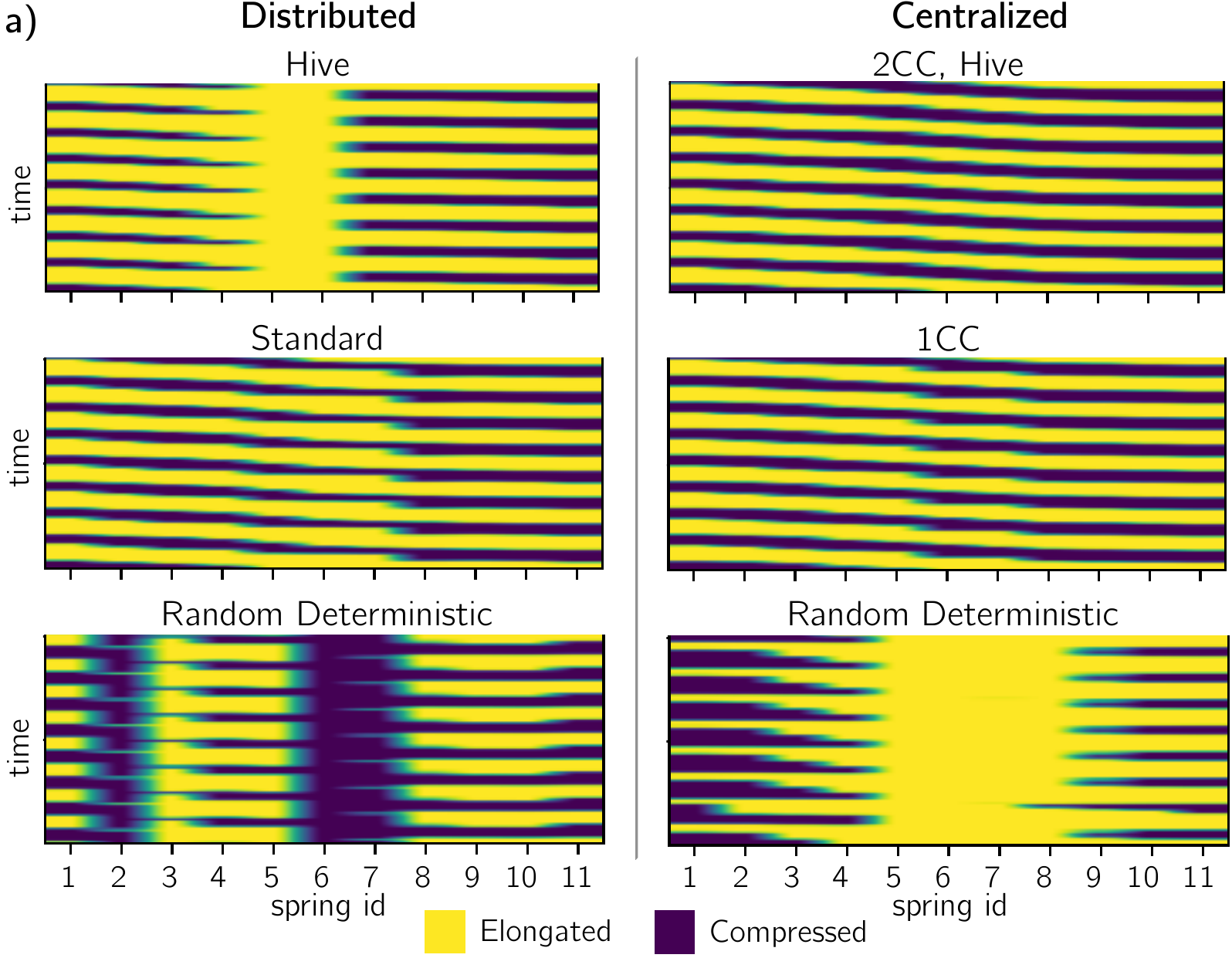}\\
    \end{subfigure}
    \begin{subfigure}[t]{\linewidth}
        \includegraphics[width =\linewidth]{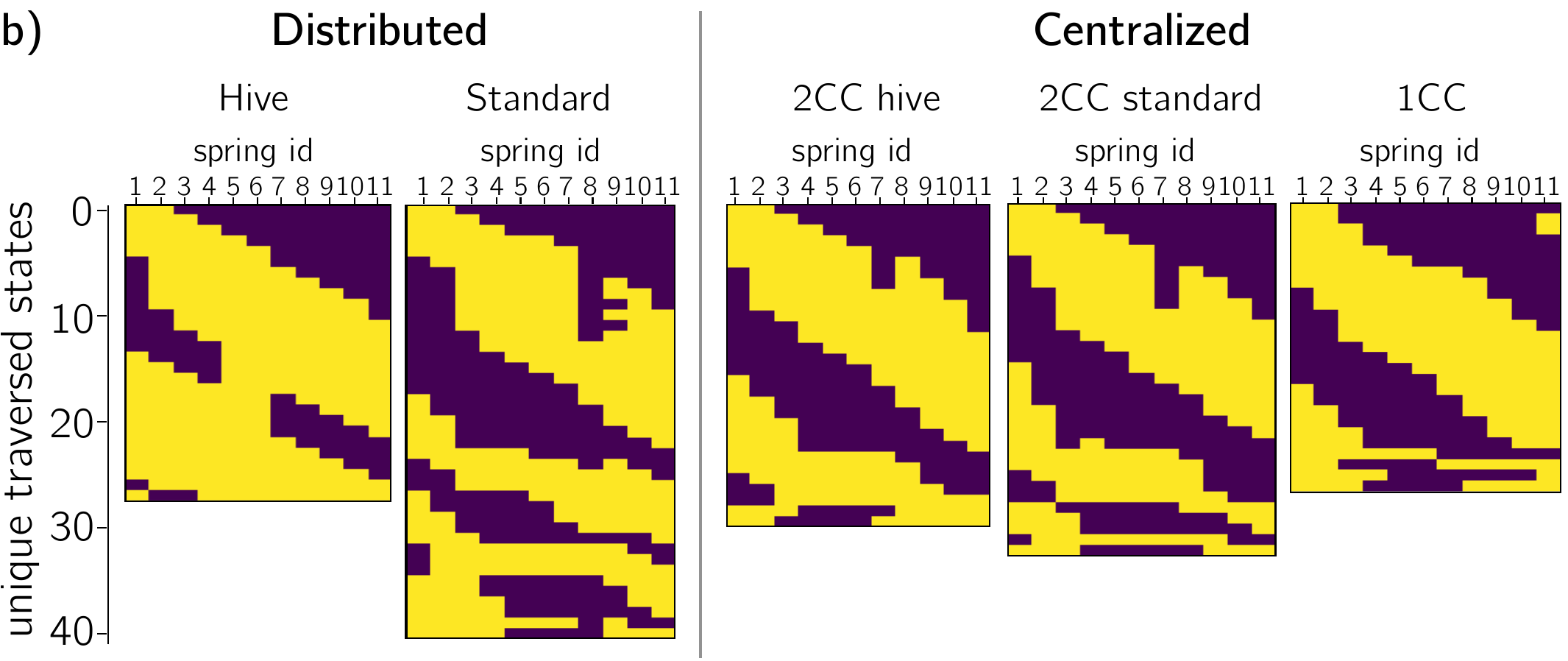}
    \end{subfigure}
    \caption{Spring states observed when playing the best policy for the different architectures considered in the 12 suckers crawler, compared to a random deterministic policy.
    a) Kymographs of states experienced by each spring (x axis) during integration: each row corresponds to a different time; time increases on the y axis from top to bottom.
    b) Unique binary states of compression/elongation of the 11 springs experienced as the crawler adopts specific policies. {
    States appear in an arbitrary order that 
    does not account for the time spent in each state nor the temporal order these states are encountered (e.g.~state 0 is not visited more often than state 1, nor it is visited right before state 1).
    }
    \label{fig:bands}}
\end{figure}

\subsection{Orchestrating a traveling wave}
The results described so far suggest that centralized control achieves better performance at the cost of exploring a higher dimensional space of possible states and actions. Why is centralization beneficial? 
Our crawler is composed of individual suckers that only have the ability to sense a binary compressed/elongated state of their adjacent springs. This makes for a system of poor proprioceptors that do not have access to e.g.~their spatial location nor to a clock. 
We find that by orchestrating the action of many poor proprioceptors, centralized control is able to ride the traveling wave generated by the CPG, despite the lack of a direct measure of space or time. 
To characterize this intuition, we 
visualize kymographs of the state of compression of the springs generated by the dynamics as a crawling policy is enacted (\cref{fig:bands}(a)). Note that the state of compression of the springs is not deterministically controlled by the adhesion pattern of the suckers.  
We find that all optimal policies generate periodic states of compression as seen by repeating patterns 
representing traveling waves of compression with speed $\approx 0.19108$, perfectly matching the underlying central pattern generator phase velocity, $v_p/x_0 = \omega/k \approx 0.19099$, with $\omega = 0.1,\,\,k=2\pi/N_s$ and $N_s=12$. This is in contrast to random deterministic policies, which generate random states of compression of the springs and cannot reproduce the CPG wave (bottom panel of \cref{fig:bands}(a)).  
Interestingly, for distributed architectures 
in the hive case the bands abruptly end near the center of the crawler, signifying that when the compression wave hits springs 5 and 6 it stops: 
a clump of springs near the center of the crawler remains constantly elongated. In contrast, for centralized architectures the wave never stops but rather travels along the crawler all the way to the end, so that the CPG movement is effectively transferred to the crawling.
Thus by orchestrating adhesion of multiple suckers at a time, centralized architectures realize a smoother traveling wave of compression. 
{ Consistent with the observation that centralized policies generate a smooth travelling wave of compression, the dynamics does not generate all possible combinations of compression and elongation of all springs, which would be $2^{N_s-1}$. Instead, }
only a limited number of compression states is visited. {More precisely, }for the optimal fully centralized policy (1CC), the 12 springs experience only 26 states of compression, about $1\%$ of the available $2^{N_s-1}$ combinations of binary states associated to the $N_s-1$ springs. { When the crawler enacts any other policy, the coupled dynamics generates more states, indicative of stumbling blocks in wave propagation. This is despite the fact that centralized policies respond to the global state of compression of all springs simultaneously, and are thus much more complex. Thus, somewhat counter intuitively, more complex policies with more degrees of freedom generate fewer states than simpler policies with fewer degrees of freedom.} 

\section{Discussion and perspectives}
Here we presented a toy model of a one dimensional crawler composed of blocks connected by springs. Each block is a rudimentary actuator representing a sucker making a single binary decision at each time: adhere \emph{vs} not-adhere. The suckers are also rudimentary proprioceptors, as they measure  the binary state of compression \emph{vs} elongation of their adjacent springs. Positional information and speed of the individual suckers is not measured nor inferred. We assume that a central pattern generator, or CPG, periodically modifies the rest length of the springs, according to a traveling wave.  {The CPG is dictated by an internal clock, independent from proprioception. This may be a useful model for the octopus, where proprioceptive feedback has been proposed to be implemented with an embodied mechanism rather than a neural process based on somatotopy~\cite{hochner2023embodied}.}   Using this toy model we have demonstrated that this group of rudimentary proprioceptors/actuators can learn to interact with the substrate and exploit the wave generated by the CPG to achieve a net unidirectional motion of the crawler. Particularly, we focused on the pros and cons of distributed \emph{vs} centralized forms of learning and control. \\
Using tabular Q-learning, we found that locomotion can be learned in a pure multiagent setting, where each sucker is considered an independent agent and with no explicit communication among agents, consistent with previous literature \cite{Hartl2025}. {Although the agents do not communicate explicitly, the result of their actions is indirectly coupled via the reward. Indeed, the reward is the net translocation which results from the dynamics of the system of oscillators, hence to maximize reward agents effectively learn how to act cooperatively. }
A purely distributed form of learning and control is cost effective in that it requires no communication and thus scales well with the number of agents and the degrees of freedom in the system.

 We find that introducing a certain degree of centralization, in the form of Control Centers encompassing several suckers, provides two benefits. On the one hand, the increased complexity of centralized policies can be effectively exploited to couple the actions of many suckers at a time. These long range correlations enable centralized policies to more effectively ride the wave of compression provided by the CPG, and ultimately yield faster crawling. On the other hand, as multiple suckers are activated at a time, failure of an individual sucker results in less dramatic loss of performance, making centralized policies 
{more robust to failure.}
 \\
Centralization of course comes at the cost of increased computational cost and requires two-way communication of proprioceptive input from the suckers to the control center, and of motor control from the control center to the suckers. We find that intermediate architectures where multiple control centers orchestrate the actions of a subset of suckers achieve the best of both worlds, providing virtually identical benefits of full centralization, but limiting the number of degrees of freedom to a manageable size. 
Ultimately, partial or full centralization overcome the limits of binary proprioception and activation by realizing smooth and continuous dynamics, despite binary input and output. 
Further work is needed to establish the optimal degree of centralization, emerging from the trade-offs between robustness, speed and computational cost, which will be relevant to understand the control architecture of biological crawlers and to design efficient robotic crawlers (see e.g.~\cite{Chen2020,Sato2010,mazzolai1,mazzolai2}).

Although the physics underlying crawling is clearly deterministic, the learning task is stochastic due to partial observability. Indeed, the deterministic dynamics of the crawler is dictated by the instantaneous positions and speeds of all suckers and rest length of all springs, but these are unknown to the agent. On top of this, in all but the fully centralized architecture, the individual agents are unaware of other agent's actions and states of compression, which yields another source of stochasticity. 
In the context of partial observability, stochastic policies may provide better performance (e.g.~adopting policy gradient methods) 
\cite{Sutton2020,Peters2008}. 
More refined forms of Q-learning may also be beneficial here, for example double Q-learning which was reported to be more robust when dealing with stochastic Markov Decision Processes~\cite{Hasselt2010} or other modern RL-techniques, for instance based on Deep-RL which couples Neural Networks to Q-learning~\cite{Sutton2020,Plaat2022,Kim2022}. Some of these techniques may prove more effective than our simple approach based on tabular Q-learning. However they remain outside the scope of the present paper where we focused on disentangling the benefits and limits of distributed \emph{vs} centralized learning architectures. For example, a complex neural network naturally couples inputs and outputs and may thus not be the natural choice to parse the role of centralization vs distribution. 

In contrast, previous biologically inspired work {with simple models of crawling (or templates~\cite{Full})} focused on either a fully centralized~\cite{Mishra2020} or fully distributed (hive) control~\cite{Hartl2025}. In~\cite{Mishra2020}, a biologically plausible model of crawling is proposed 
which also uses rudimentary forms of  proprioception and control: input is limited to the identity of the most compressed body segment and actions correspond the designation of the sole active neuron in the neural network controlling muscular contraction, while adhesion is not controlled. 
The authors show with a fully centralized Q--learning architecture that the central pattern generator can be learned, rather than imposed as we do here. The benefits of distributed \emph{vs} centralized control {for animal locomotion} have been studied in several other contexts { and  algorithms, and generally pertain added stability in uneven terrain, adaptation, versatility, see e.g.~\cite{yasui2019decoding,yasui2025multisensory,suzuki2021spontaneous,neveln2019information}.}\\
In~\cite{Hartl2025}, swimming of an agent is achieved through training of a genetic algorithm with a fully distributed architecture, where individual actuators are all forced to learn the same policy -- which we have called hive update here. Again, the actuators are not suckers but rather beads which control the restoring force between their neighbors. The hive update was found to generalize well to a different number of beads, with no need for re-training. This is consistent with our finding, where we also observe that the hive update almost invariably learns the same policy, independent on the number of suckers composing the crawler. However, while we focus on binary sensing and control, Ref.~\cite{Hartl2025} considers actuators with full proprioception, including the position and speeds of their neighbors. 
An advantage of considering binary proprioception is its simple implementation and interpretation. For example, our hive update can be understood in simple terms: suckers learn to only adhere when they are pushed and pulled backward from both sides. 
 
Here we have focused on the learning process that enables suckers to achieve crawling by combining mechanical contractions and interaction with the substrate. 
In our model, thrust is achieved by letting suckers control patterns of adhesion to the solid substrate, whereas contraction results from the stereotypic wave of compression and the dynamics. 
In refs.~\cite{Mishra2020,Hartl2025} switch the controls: the actuators learn patterns of contraction, and interaction with their solid and fluid environment respectively is embedded in the dynamics. 
Finally, in bio-inspired models of sea star locomotion~\cite{Heydari2020,Po2024}, thrust is achieved by fully prescribing --rather than learning-- actions. Synchronization here emerges through the dynamical interaction of the substrate with the tube feet, which are physically connected to the body, with a mechanism reminiscent of the Kuramoto model of synchronization~\cite{Kuramoto1975,Strogatz2000}.\\

Here we parsed the pros and cons of distribution \emph{vs} centralization for learning the internal coordination among body parts, aimed at translocation. 
The benefits of centralization for speed and robustness point to a potential origin of selective pressure towards more centralized forms of nervous systems. To what extent early nervous systems allowing the Cambrian explosion may have emerged from the need for internal coordination \emph{vs} from the need of processing complex sensory information is currently debated, see e.g.~\cite{Jekely2015,Arendt2016,Keijzer2017}. 
Our results are dictated purely by the need to internally coordinate body parts to generate thrust, but the role of centralization may become even more prominent when sensation and locomotion are coupled to aim toward a specific target. Indeed, this requires that each unit senses not only the internal state of compression of the organism, but also sensory signals from the environment, from e.g.~vision, audition, taste and olfaction. Our approach may be further extended to sensory navigation, by enriching the state definition with sensory cues from the environment and rewarding not only speed but also the successful reaching of a sensory target. \\
Our model is inspired by octopuses and other cephalopods, organisms with a distributed nervous system whose proprioception remains elusive~\cite{Wells1960,Gutnick2011}. 
Control centers coordinating adhesion of a few suckers with rudimentary proprioception achieves fast translocation without incurring into the computational cost of a fully centralized brain. 
Our approach may be further developed to quantify performance, robustness and computational cost as a function of the number of control centers, to select an ``optimal'' number of control centers for crawling, aimed at both fast locomotion as well as sensory navigation toward a target. Control centers may be adapted to model how ganglia distributed along the arms of an octopus control locomotion. 
To what extent the robustness, performance and computational cost of crawling shape the connectivity of the ganglia distributed along the octopus arms' and the organization of their taste-by-touch sensory system~\cite{Bagheri2020,Allard2023,VanGiesen2020} is an exciting avenue for further research.

\section*{Acknowledgments}
This research was supported by grants to Agnese Seminara from the European Research Council (ERC) under the European Union’s Horizon 2020 research and innovation programme (grant agreement No 101002724 RIDING) and the National Institutes of Health (NIH) under award number R01DC018789. This work represents only the view of the authors. The European Commission and the other funding agencies are not responsible for any use that may be made of the information it contains.
\newpage
\appendix

\section{Validation}
\subsection{Periodic crawler}
\label{appendix:analytic}
\begin{figure}[hb]
    \centering
    \includegraphics[width=\linewidth]{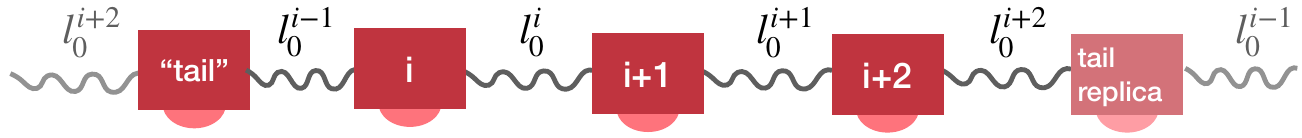}
    \caption{Sketch of the periodic crawler model.}
    \label{fig:appendix_sketchPeriodicT}
\end{figure}

\begin{figure}[ht]
    \centering
    \includegraphics[width=\linewidth]{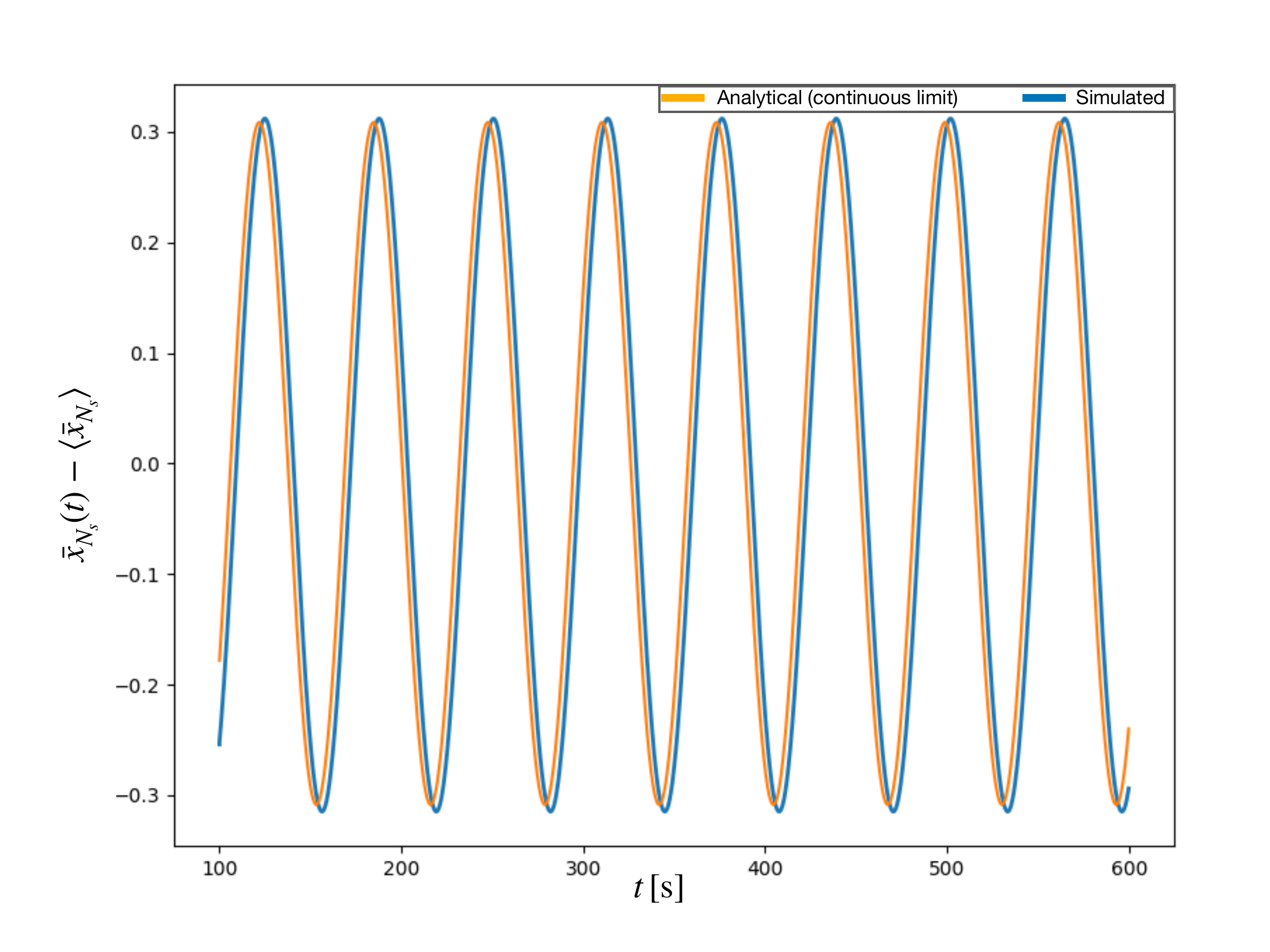}
    \caption{For validation purposes, we here track the movement of the head of a 10 suckers crawler in overdamped dynamics and with periodic boundary conditions (infinite crawler ideal model) and compare it to the analytical solution of \cref{eq:unperturbed}. The position are reported in the reference frame of the time-averaged position and in normalized units $\bar{x} = x/x_0$. The amplitude of the oscillation is here $a=x_0/3$.}
    \label{fig_appendix:validation_periodic}
\end{figure}
 
\begin{figure}
    \centering
    \begin{subfigure}[t]{0.5\textwidth}
        \includegraphics[width=\linewidth]{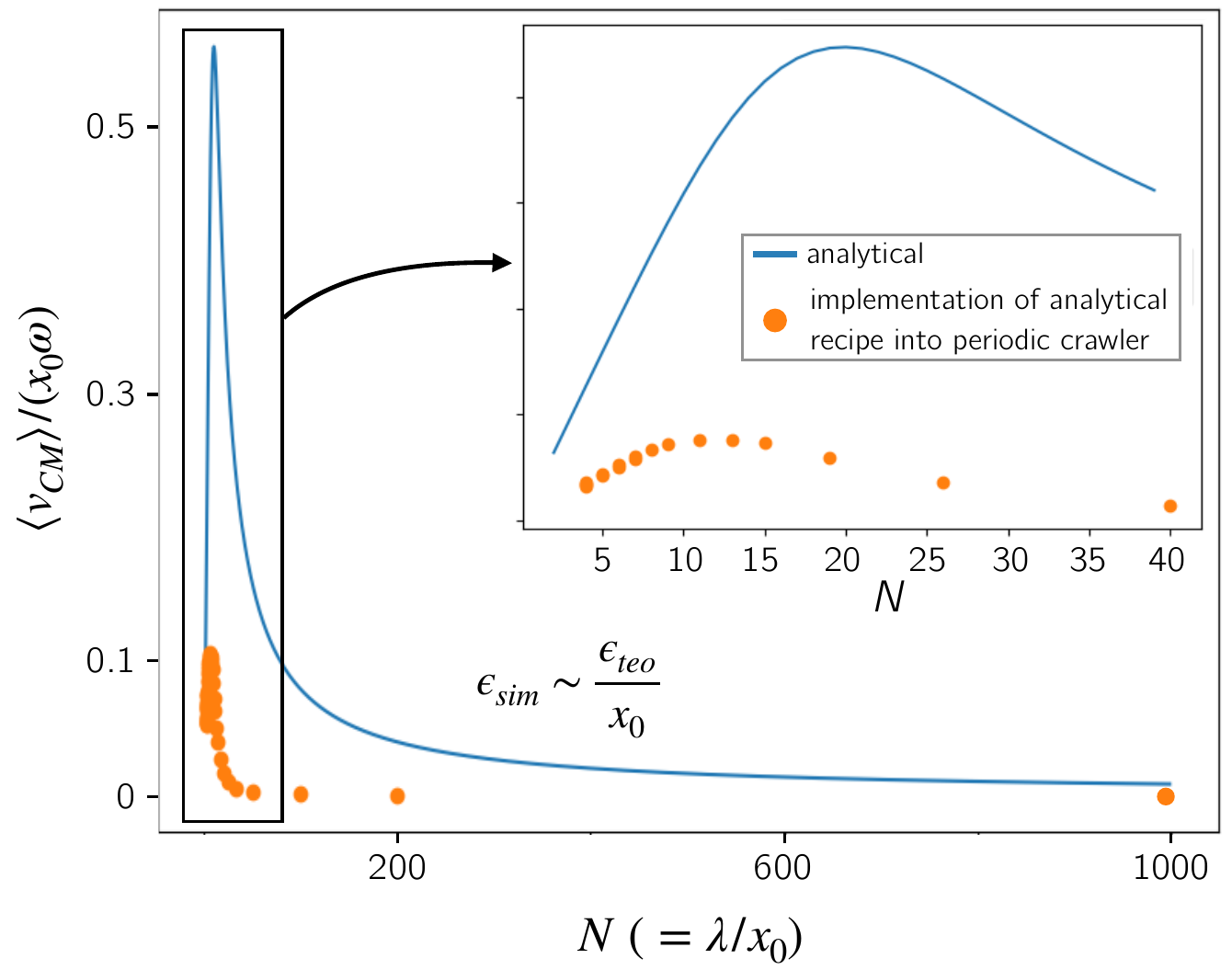}
    \end{subfigure}
    \begin{subfigure}[t]{0.5\textwidth}
        \includegraphics[width=\linewidth]{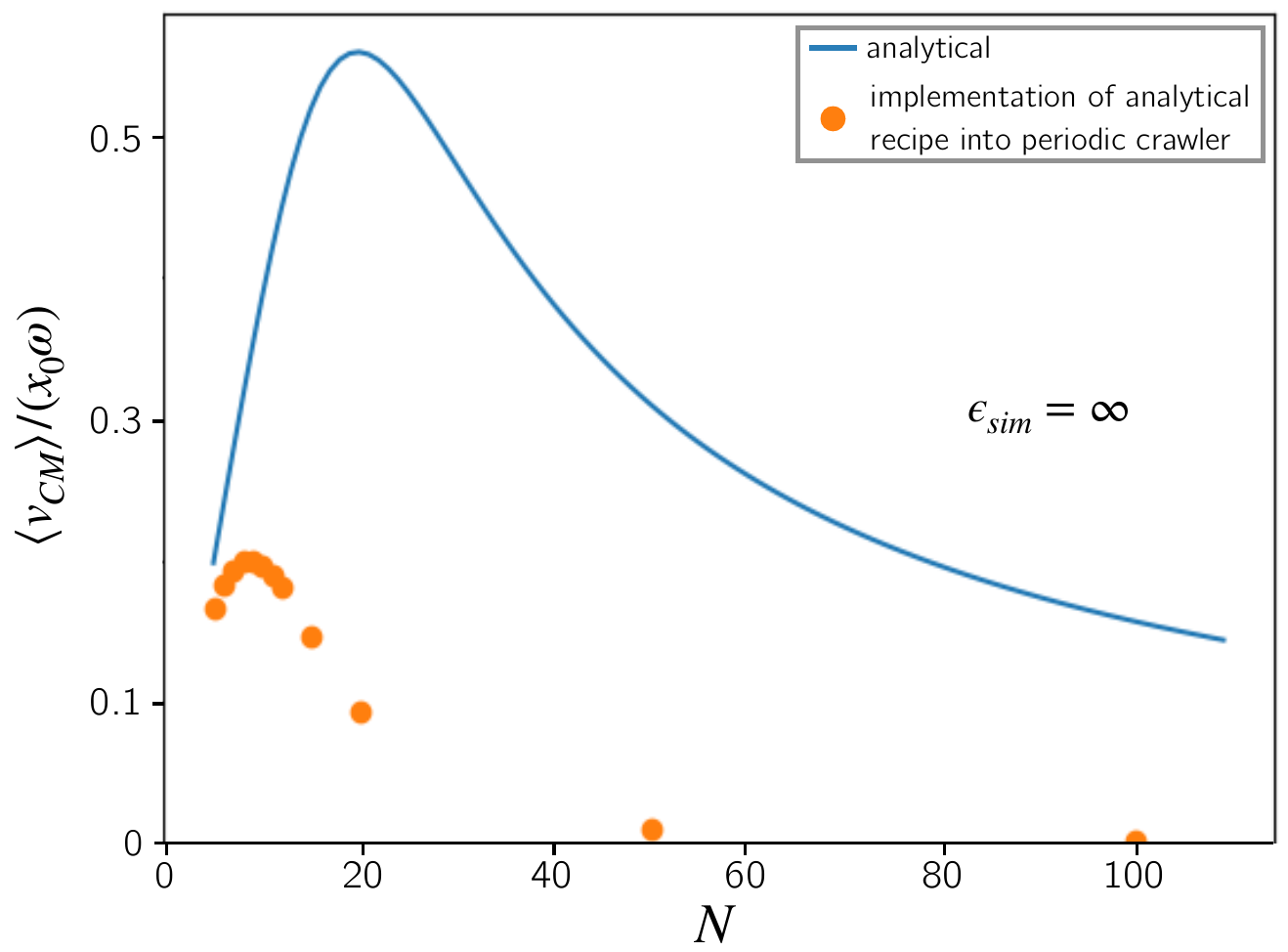}
    \end{subfigure}
    \caption{Comparison between eq.~4.9 of \cite{Tanaka2012} 
    (blue lines), and the implementation of \cref{eq:anchoring_pulse} in the simulator (orange dots) for the periodic crawler model. 
    a) A more honest comparison with the perturbation regime used in Ref.~\cite{Tanaka2012} which accounts for the small adhesion limit, using $\epsilon = 1$ in the simulator.
    b) With infinite adhesion in the simulator as the one used generally in this work. 
    }
    \label{fig:appendix_analytical}
\end{figure}

As discussed in \cite{Tanaka2012}, in the continuous limit the equilibrium spring position (central pattern generator) is $l (s,t) = x_0+ a \sin(\omega t -k s)$, with $s = n x_0$, $k_c = 2\pi/\lambda$ and $\lambda = Nx_0$ with $N$ the periodicity.
It can be proven \cite{Tanaka2012} that without any perturbation, i.e.~no adhesion, that is uniform and constant friction across all the crawler, $\zeta_0$,  the continuous equation of motion for a sucker at location $s$ is
\begin{equation}
\label{eq:unperturbed}
    u^0(s,t) = C \cos (\omega t - k_cs -\theta)
\end{equation}
with
 $\theta = \tan^{-1}\tilde{\omega}$, $\tilde{\omega} =\zeta_0\omega/(k^2\kappa)$, with  $\kappa$ the springs elastic constant, and $C = a/N \cos \theta$.
We can approach the system described in \cite{Tanaka2012}, where they assume an infinite crawler, by considering a discrete crawler with periodic boundary conditions and check how the unperturbed solution (no sucker adhering) compares to the analytical prediction of \cref{eq:unperturbed}.
As illustrated in \cref{fig:appendix_sketchPeriodicT}, periodic boundary conditions are implemented by considering a virtual replica of the tail connected to the last right-hand sucker of the crawler.
The comparison is made in \cref{fig_appendix:validation_periodic} where we track the head position of the simulated periodic crawler with no adhesion (so no net movement of the center of mass, $v_{CM} = 0$). The comparison is extremely good with what predicted from theory.

When the crawler is perturbed by the adhesion pulse of \cref{eq:anchoring_pulse}, { an asymptotic solution is available from } ref.~\cite{Tanaka2012} (\cref{fig:appendix_analytical}).
The implementation of the optimal crawling strategy (\cref{eq:anchoring_pulse}) in the periodic crawler  discussed here, 
{has the same qualitative behavior predicted by the theory, \cref{eq:vmax}, with a different location and magnitude of the maximum. These discrepancies likely arise due to our approximation of the infinite crawler with the periodic crawler.} 

However, as also shown in the Results section for the finite crawler, we have a fairly good qualitative comparison with the theoretical behavior, namely the existence of a maximum in $N$ ($=N_s$, the number of suckers for our parametrization), and a decay for $N\rightarrow \infty$. The position of the maximum is not exactly the one predicted by the analytical theory but similar (with $\omega = 0.1$, $N_{\max}^{finite} \sim 12 <\tilde{N}_{\max} = (2\pi)/\sqrt{\omega} = 20$).
Interestingly, as illustrated by \cref{fig:results} in the main text, the functional shape of the dependency with $N_s$ is preserved even when crawling policies allow for the adhesion of multiple suckers at a time which was excluded from the analytical treatment of Ref.~\cite{Tanaka2012}. 

\subsection*{Damped limit}
\label{appendix:damped}
For validation purposes, we can drop the overdamped assumption and solve the full damped equation for the finite crawler \cref{eq:springs}.
The overdamped regime is already approached when {inertia balances the restoring force of the spring, i.e.~$m\omega^2/\kappa=1$}.
Preliminary results also suggest that policies learned in the damped regime are not significantly different from the overdamped ones. Therefore, we expect that the conclusions reached in this work should still hold in the damped scenario. However, further verifications would be needed to assert this claim unequivocally.

\section{Effect of CPG periodicity over sucker importance}
\label{appendix:periodicity}
\begin{figure}
    \centering
    \includegraphics[width=\linewidth]{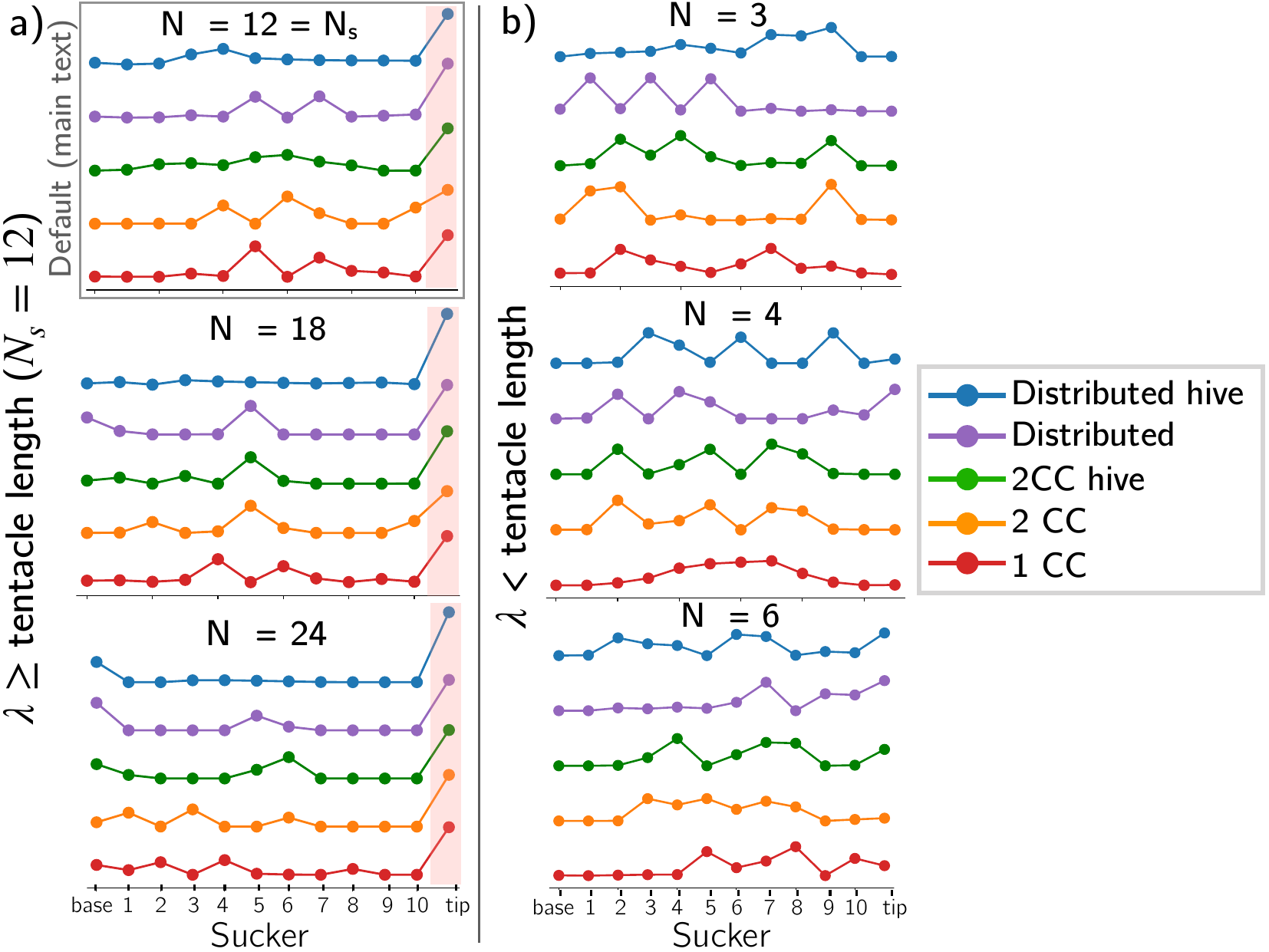}
    \caption{Effect of different wave forms of the CPG over the first order sucker importance for a $N_s = 12$ suckers crawler. We can note that when the wavelength is equal or larger than the crawler length, $N\geq N_s$, the head is always the most impactfull sucker for all the considered learning architectures. This invariance does not hold anymore when $\lambda<L$.}
    \label{fig:appendix_periodicity}
\end{figure}
In this work in general we assume for simplicity that a single wavelength is contained in the crawler, $\lambda = L$ (that is, $N=N_s$).
As discussed in the main text, we found that when playing a crawler policy the head is on average the most impactfull sucker upon failures, whatever the considered learning architecture.
To establish if this is a general property and how it correlates to the physics of the system, we analyzed for the 12 suckers system how this property is affected when dropping the assumption that the CPG wavelength, $\lambda$, is identical to the crawler length, $L$. 
The results, shown in \cref{fig:appendix_periodicity}, indicate that the head is always the most determinant sucker only when the CPG wavelength is $\lambda\geq L$. However, when $\lambda<L$ we cannot find a single sucker to be the most important for all the considered architecture, but its identity varies with the considered policy (architecture) and wavelength.
We do not have yet a definite explanation of this phenomenon, but this results already indicates a non-trivial interplay between the underlying physic of crawling, namely the (discrete) placement of each sucker with respect to the CPG, and crawling policies.

\clearpage

\bibliographystyle{apsrev.bst}

\bibliography{crawlerBiblioNew}

\end{document}